\documentstyle[12pt,a4]{article}

\begin{document}

\title{ \vglue-3cm
\hfill {\normalsize HEP-TH 9410085\\
\hfill with  additional remarks}
\vskip 3cm
\Large\bf Reminiscences about Many Pitfalls and Some Successes of QFT
Within the Last Three Decades}

\date{September 1994}

\author{\bf B. Schroer\\
Freie Universit\"at Berlin\\
Institut f\"ur Theoretische Physik\\
Arnimallee 14
14195 Berlin}
\maketitle
\centerline{To appear in Reviews in Mathematical Physics}

\newpage
\section {Introductory Remarks}

Laymen and sometimes even physicists think of natural sciences, in particular
of theoretical and mathematical physics often as subjects, which unfold
according to an intrinsic logical pattern, with the limitations being set only by the conceptual and
(in case of mathematical physics) mathematical developments of the times.
This view certainly cannot be maintained in view of the present stagnation and
crisis which in particular affects  QFT, an area which in the past has been most
innovating and fruitful, also in relation to other important areas of 
theoretical physics.

In this article I try to analyse this situation using my personal experience
of 30 years of scientific carrier life. One of course always hopes that by doing this one
may recover physically fruitful ideas which either got lost completely or at
least to the younger generation. In good times this may be only of historical
interest, however in times of stagnation and crisis one may expect that the 
past
may suggest new avenues into the future of new concepts and principles in
physics (and not just as a service to produce new mathematics as it seems to
be presently). 
I want to emphasize that in writing these notes, I followed some natural current
of recollections (written within three weeks, during a serious health crisis)
and I certainly did not attempt to give a complete account of all contribution
which had an influence on my work  and on my thoughts. So e.g.,
the analytic work on S-matrix properties (Lehmann, Oehme, Martin,...), the work of
Goddard, Kent, Olive as well as Nahm on algebraic conformal QFT and also
 some recent and not
yet fully published work of Longo, Rehren, Roberts and also Wassermann
(investigations of the field algebra -- observable algebra connection by using
Jones subfactor theory) are not mentioned at all. 
Recently I also became
aware of promising attempts of using chiral conformal QFT for semi-phenomenological
analysis in condensed matter physics (edge states in layered quantum structures).
But since I was unable to relate the interesting work of Afflek, Fradkin,
Fr\"ohlich, Laughlin, Ludwig, Wilszek and many others to my idea of 
universality classes of quasi-particles (see the last section) in any
convincing way, also that work will go uncommented.
Even contributions with some own involvement as the old investigations
of ``semi-nonperturbative" (abstraction from every order of renormalized
perturbation theory) ``soft" versus ``hard" symmetry-breaking of Symanzik,
Coleman, Stora, Ben Lee, Jackiw,  and others or the recent partially successful attempt to
understand space-time covariance from raw local quantum physics (using the Tomita-Takesaki
modular theory) by Borchers and Wiesbrock will go unmentioned in the text.
 This would have overburdened
these notes and diminished their spontaneity. 

Last but not least this would also
have created an aura of a ``wizzard on everything". I am certainly not even close
to anything like this. My understanding of recent geometric developments and
string theory is flimsy, and even in such physically important areas
as the GSW semi-phenomenological theory of electroweak interaction (and its extensions),
I do not really feel at home (admiration about achievements is not enough!)

After a very long absence from algebraic QFT, I returned to this area
(like the prodigeous son in the bible) around 1987, naturally beeing very impressed
by its gains since 1968. In such a spirit of enthusiasm, one is inclined to
proselyte others. I was not successful in that, perhaps with one minor
exception. In 1988 at the XVII International Conference on Differential Geometric
Methods in Theoretical Physics, I met Ivan Cherednik and he seemed to like
those ideas. We emptied a bottle of russian vodka together (from which also
Ruth Lawrence took a very well-supervised sip).

In these notes, I use the terminology ``algebraic QFT" in a bit more general
sense than that of ``net-theory" (with the exception of the last section).
I call ``algebraic"  everything which is not obtained by functional integrals
or perturbation theory. I also take the liberty of not giving references to
those papers which are common knowledge in physics or mathematics (thus protecting
these notes against any non-intended misuse).

The speed of progress in theoretical physics always has been more determined
by discoveries than by inventions. There are two types of theoretical discoveries,
those which are directly motivated by and linked to an experimental discovery
(often with considerable hindsight and guesswork) and there are also purely
conceptual discoveries in which confrontation with known principles and resolutions
of apparent paradoxa play the prime role. The latter type of discoveries are
often made by physicist with a considerable mathematical physics background
i.e. physicist who know to use the art of mathematics for conceptual precision
in physics. Certainly the early development of Q.T. as a description of atomic
spectra belongs to the first category. 

One is also inclined to count the
discovery of QED and its renormalization into this category, although the
necessity of quantization of the Faraday-Maxwell field as a matter of 
consistency with the principles of Q.T. of charged particles by Bohr and
Rosenfeld has a strong element of a pure theoretical discovery. Most of
Einsteins discoveries (special and general relativity) consisted in extracting
new principles from already existing formulas or equations. The Kramers-Kronig
dispersion relation which dominated particle physics in the later 50$^s$
is another example of a pure theoretical discovery: it was obtained by
investigating the consequences of Einstein causality within relativistic
scattering theory. The importance of its subsequent experimental verification
in forward high-energy nucleon-nucleon scattering cannot be overestimated.
Our confidence that causality and locality hold up to present day energies
(or short distances) is founded on these verifications. The protagonists
of this modest (in its physical aims) but nevertheless important discovery
should not be blamed for the fate which these ideas suffered afterwards
when more ambitious people tried to convert them into the first
\footnote{The first TOE I met in my life was however not this but
(during school-time in the GDR) rather Marxism-Leninism.} ``TOE"
(to be more precise, a theory of everything minus quantum gravity). 

It is
interesting to observe, that the mechanism according to which ``TOE's" are invented
always seem to follow a similar pattern. One formulates formal as well as often
physically reasonable requirements (the latter came from QFT) in such a
way that one encounters a highly nonlinear and extremely uncontrollable
(both in a mathematical and physical intuitive sense) situation. Usually such
structures resemble the infinite nonlinear system of the Schwinger-Dyson
equation in QFT. Although one has no solution at all, one hopes (or dreams) that it
has a unique solution. After many years of thinking one finds that such a
``bootstrap" situation has myriads of solutions, usually by ``defusing the
nonlinear dynamite" via additional physically motivated ideas leading to a
linearization.

In the case of the S-matrix bootstrap for two dimensional QFT the key idea
which linearized the ``nonlinear dynamite" was the factorization equation
which characterizes the family of two-dimensional integrable local QFT's.
We will cast some additional light on this interesting (not only historically)
situation from a more modern point of view later on.

Two important yet even more modest discoveries of the last decades with
experimental verifications (and in the second case even additional consequences)
are the Bell inequalities and their maximal violation and the appearance of
Berry's phase in different areas of atomic, nuclear and optical physics.

The issue of QCD and more generally of nonabelian gauge theories in the
context of discoveries versus inventions is somewhat complex. On the one hand
there is the minimal interaction recipe and the quantization rules by which
the classically crystal-clear Maxwell-like theory of fibre bundles is supposed
to be converted into a consistent QFT whose intrinsic quantum content remain
somewhat hidden (and is only expected to show up after not yet known laborious
nonperturbative calculations or by Monte Carlo simulations of lattice systems
which are expected to lie in the same critical universality class). This has
clearly more the ingredients of an invention. On the other hand the observation
that non-abelian gauge theories in the perturbative region are among all local
renormalizable Lagrangian field theories the only ones (with some restriction
on the number of matter fields) which exhibit the experimentally relevent
phenomenon of asymptotic freedom is clearly a purely theoretical discovery.
If one could derive it from a renormalization group equation not containing a
gauge- and renormalization scheme dependent coupling constant, but rather a
physical mass ratio (and therefore may also have an intrinsic meaning in the
nonperturbative regime) then the asymptotic freedom statement would even
get the status of a structural theorem like the nonperturbative version of the
Nambu-Goldstone-theorem or the ``free charge versus screening" theorem of abelian
gauge theories.

Last but not least, the elaboration of the unified perturbative electro-weak
theory and it numerous experimental verifications makes it the most successful
phenomenological discovery of all times up till now. But to transform this scheme
into a fundamental theory with clear principles and fewer parameters has 
turned out to be difficult indeed. One gets more and more the impression that
the prize for that instant phenomenological success was 
(like in Greek mythology)
the entering of a labyrinth from which there is no easy way out. 

In fact the
theoretical situation has become so complicated that even formally motivated
proposals coming from mathematicians (as ``non-commutative geometry") would
be accepted gratefully if they only lead to simplifying, parameter-reducing
prescriptions (like e.g. the minimal gauge coupling rule).

Inventions which either lack an experimental basis or are not asked for
by known theoretical principles (as a string theory of everything, or the
more recent formal $q$-deformations) presently seem to enjoy a large amount of
popularity. It is my conviction that this is more related to the
sociology of contemporary theoretical physicist and their rapid electronic
communications than to the actual physical content of these inventions.
 To be sure, inventions
in earlier times were as copious as they are nowadays. But as a result of a
smaller number of protagonists and researchers, and also as a result of lesser
sophistication of their mathematical description, they had a greater chance to
fade away  by natural death in case they did not work in physics.

However, having said this, I do not want to be misunderstood to nourish
nostalgia for a mathematical stone age in physics or any restrictions to
pedestrian methods.

But sociology in theoretical physics has changed definitely from what it
used to be. When Einstein together with just one or two collaborators
worked on the aborted unified theory, there was no lack of courage to
tell them that they were on the wrong physical track.

\section{ The Early Days of Algebraic QFT}
Already very early in the history of QFT, there was the desire to obtain more
insights by involving pure quantum principles without any classical parallelism
referred to as ``quantization".
\footnote{As Haag pointed out in his book [1], Bohr's famous remark on the
classical versus quantum relation (``we must be able to tell...")
has often been misinterpreted. Bohr's correspondence principle, that
quasi-classical structures appear under very special circumstances,
does not constitute a free ride for ``quantization" either.}
 The most successfull early attempt was that of
Wigner who worked out a complete classification scheme of relativistic 
one-particle states by purely group theoretical means, [1] i.e. by classifying 
irreducible representation of the proper orthochronous Poincar\'e group and later
even
including the various reflections. The methods were those of ``induced
representations" (the ``little group"-method) going back to Frobenius and
later perfected by Mackey. The method immediatly led to local free field
equations, and it was very easy to decide when two apparently 
different-looking system of 
linear field equations (say obtained by the quantization method)
are equivalent. However in spite of Wigner's work, only in the early fifties 
the senseless mass production of new higher spin field equations subsided 
when physicists began to appreciate the significance of Wigner's work and the relevance of 
Fock space. This work together with the seminal paper on superselection rules
(more specifically the univalence superselection rule between half-integer
and integer spin states) by Wick, Wightman and Wigner [1], set the stage for the
two important formulations of framework for nonperturbative QFT:
Wightman's theory [2] (with important contributions of Lehmann Symanik, Zimmermann,
Glaser and Nishijima) and the Haag-Kastler theory (with important contributions
of Araki and Borchers). Whereas the first started by ``axiomatizing" physical and
technical properties of local covariant fields (with LSZ scattering theory
giving the bridge to Wigner's one particle states and their multiparticle
extension by free fields in Fock space), the latter vastly extended the 
W-W-W-superselection idea to generalized charges and viewed QFT as an
operator-representation-theory of nets of observable algebras, thus 
emphasizing the locality principles known from the classical Faraday-Maxwell
theory ab initio (i.e. not through quantization) in an algebraic (von Neumann
algebras, $C^*$-algebras) setting.

``Axiomatic" QFT was, at least at the beginning, just a pragmatic compilation
of all those physically motivated properties which, at that time, were
susceptible to a reasonably clear mathematical formulation (but without
caring too much about the possible interdependences) . The word
``axiomatic" attributed to this list an aura of permanence and mathematical
and physical stability which was somewhat detrimental. The more recent 
terminology ``local quantum physics" [1] exposes its aims more clearly:
incorporation of the locality principles of Faraday-Maxwell-Einstein into
quantum physics with the least amount of additional inventions. The global
structures, pointing into the direction of global geometry and topology, are to be
achieved at the end, and do not belong to the foundations.

At a conference in the early 60$^s$, Goldberger (a leading expert on the
phenomenological use of dispersion relations at that time) once said:
``the contribution of axiomatic QFT to physics has been smaller than any
preassigned $\varepsilon$." In retrospect one notes that only those parts
of the dispersion theory survived (and are still tought in courses) 
which were reasonably close to the principles
of QFT. 

The detrimental and prejudical effect of that word ``axiomatic"
is to a large
degree responsible for the fact that deep and interesting results of general QFT went 
unnoticed or got willfully ignored, even if they were very relevant to the
main-stream physics (the reader will come across a lot of instances of this
kind in the subsequent sections). 

But this only explains the situation up to
1980. The lack of interest in pure local quantum physics after 1980 is
definitely a result of the exagerated expectations in the importance of
geometric and global topological structures for the formulation of quantum
physics. Now, after this development has, to some degree, returned
mathematics, algebraic QFT may have the chance to fill the void left behind.

Having indicated in the introduction that I will keep my thoughts
free-floating and uncensored, let me add the following. While at the University
of Illinois, I once read in the newspaper about the failure of a small
tomato-ketchup producer. He had the unpopular (as it later turned out)
idea of producing a pure, strong-tasting and natural ketchup from high
quality tomatoes, without the standard additives of pineapple juice and 
chemical ingredients. But most people did not want to miss those ingredients
to which they were already addicted.

One of my first contribution in collaboration with Haag [3] was a reexamination
of those postulates of QFT which were already studied in the context of
local covariant fields by Wightman, but now in the light of a more
algebraic setting. We had little problems in convincing ourselves that the observable
algebras belonging to compact space-time regions (e.g. double cones) should 
be indecomposable  factors.
\footnote{ The factorial structure and the closely related duality property
were on Haag's mind already before the start of that collaboration.
For this reason, I recently proposed to my collegues to call it
``Haag duality", a proposal which enjoyed widespread acceptance and also
facilitates the distinction to other notions of duality.}
 However when it came to question of what von Neumann-type
 these algebras belong to, we could not find an argument that they are, as in ordinary quantum
mechanics of type $I$ (although, at least at the beginning we thought that they
ought to be). The other types, especially type $III$ were already described
in the textbooks of those days (we had a German translation of  Naimarks
book on ``normed rings") but they were
 somewhat ``queer" and very rare indeed. But we were
careful enough not to let our prejudices enter our article. Some years later,
we learned through Araki's work [1], that those local algebras are type III
von Neumann algebras which, in contradistinction to standard Q.M. contain only
infinite dimensional projectors. At a summer school in Boulder, Colorado, Irving Segal
presented a proof that the local algebras are type $I$ factors. Araki, Haag
and I were in the audience. We looked at each other with a twinkle in the 
eyes and I could not supress a certain feeling of malicious pleasure.
If you almost run into a pitfall yourself, you feel less stupid if this
happens to somebody else, especially if his reputation is as high as that of
Segal. Looking at his calculations more closely, we realized that he did a 
correct calculation on 
the wrong algebras (i.e. not the physically relevant causal covariant algebras).

Such an episode on something which, from a physical point of view, appears so
esoteric, may hardly seem worthwile mentioning. Well, it is not really that
esoteric. Recently a Phys.Rev.Letter was published [4] on an apparent causality breakdown
in relativistic QFT. The main claim was, 
that Fermi's calculation on causality and his conclusion that $c$ remains the
limiting velocity (even in quantum theory) were wrong. After this article
passed the Phys.Rev.Letter referees and the issue was taken up by the
editor of Nature, there was nothing which could stop the making of a new 
hero who allegedly had shown that there is no principle which prevents ``quantum"
time machines. For the world press, this was a welcome opportunity to fill
their summer lull by something different from ``Nessy" of Loch Ness. 
For physics it was a bit of a scandal of the dimension of ``cold fusion" or the 
alleged outwitting of causality by quantum mechanical tunneling (a closer
examination may actually reveal that the recent hoax and the
tunnelling stuff can be reduced to the same conceptual flaw). 

The mistake was not computational, it was purely conceptual. The author started
 from the wrong (disproved in the mentioned work
of Araki) assumption that the local algebras of relativistic quantum field
theory have minimal projectors, (i.e. that local unrefinable observation
are possible) like those Heisenberg-Weyl algebras of ordinary Q.M..

This event, more than anything else, casts a strong light on the deep
crisis of contemporary particle physics and quantum field theory
(with increasing emphasis on the entertainment value of physics
\footnote{A good illustration is the abstract of the recent paper by R.G.$M\cup\phi$
``The fractional quantum Hall effect Chern-Simons Theory, and integral lattices".
ETH Z\"urich preprint 94/18}).
Perhaps one should consider it more as a social crisis caused by the present 
generation of theoretical physicists whose mathematical sophistication
is either limited to Q.M. or 
developped too much into the direction of differential geometry and algebraic
topology to such a degree, that important conceptual gains in local quantum
physics got lost or were not even noticed. 

To me it is somewhat sad that notions
of causality and localization of states which are at the core of QFT (and 
condensed matter physics) are so little known or appreciated by the
majority of mathematical physicists. After all, the chronology of their
evolution is the most fascinating part of nonperturbative QFT. Wigner took
his findings on the impossibility of relativistic localization within a one
particle space (the Newton Wigner localization [1]) so serious, that he used it
as a criticism against the QFT of the  50$^{s}$. 

Especially those
concepts related to localization of states on causal nets really took
some time to be conquered. In my view, the most poignant formulation is
a fairly recent one by Fredenhagen: a state $\omega$ is localized in a space time
region $O$, if outside (i.e. on local subalgebras in the causal complement $O'$) it is dominated
by the vacuum state $\omega_0$. This localization concept, which already
appears in a previous work of Buchholz, is not only stable
under composition of states (a property shared by the closely related DHR
localization), but it maintains this stability even under purification.

It should be added that the causality affair was closed (hopefully!) by a
beautiful remark of Buchholz and Yngvason [5] (also published as a Phys.Rev.Letter).
This paper, although being crystal-clear in its content, is unfortunately
very laconic on the history and the background of that issue, and therefore
its main content may get somewhat lost for nonexperts.

While still at Hamburg University, I got to know (mainly through their 
lectures) Profs. Pascual Jordan and Wilhelm Lenz. Most students know one of
Prof. Lenz's contributions via the Lenz-Runge vector in the treatment of the
integrable Kepler-2 body problem. The number of people who actually know that
he invented (and published in a one page note) the ``Ising model" in 1912
(before the Bohr atomic model) and that his student Ising solved the 
one-dimensional version in his thesis (although from his findings he drew
the wrong conclusion that the lattice version of the Boltzmann-Gibbs 
statistical mechanics seems to be incompatible with a phase transition) is quite
small.

When one enters a field as theoretical physics as a student with idealistic
principles and learns that the name of the model has little to do with its
original inventor, one is a bit shocked. Later one notices that baptizing
things by names of people is quite often unrelated to the true history of
evolving ideas but rather reflects fashions and sociology and not always 
the actual scientific content. 

Recently I came across a fascinating interview [Atiyah, collected works] which
Atiyah gave (a long time ago) to a science journalist. He mentions that the
physics Nobel prize has a bad effect on physics. I completely agree with 
that, since already since the time of Einstein, the Nobel prize has an effect of
``vaticanization" (immortality, infallability, power) in science. The Fields
medal however was thought to be different in this respect.
After such a long time a physicist  would be interested to know whether
mathematics was able to protect itself against this effect of vaticanization.

I found Wilhelm Lenz also impressive as a person. My esteem even increased
when I later got to know that he was strongly anti-Nazi. He helped and protected
Touscheck and continued to  lecture on
relativity through the second world war. For a short time during the 30$^s$,
Pauli was his assistant (at the time of discovery of the Pauli exclusion
principle). Apparently Pauli had a totally different personality from Lenz.

I decided to work with Harry Lehmann who, with the strong support of Pauli
was made the youngest physics professor at Hamburg in  succession to Lenz. 
After two years,
the attempts to learn field theory were bearing some small fruits. In my diploma
thesis I was able to characterize free fields solely by  two point functions.
The proof of this theorem was then put into a more elegant form by Jost 
(Lehmann told him about my result) and
extended by Pohlmeyer [2].  Four years later, when for a short time I shared an office
at the University of Illinois with Marc Grisaru, he asked me whether I already
know the characterization of electromagnetic free fields via a vanishing
current-two-point-function which he proved in collaboration with Federbush and
Johnson (I think it was important for Johnson in his short distance
studies of renormalized QED). Well, of course I did know about such a theorem.

I was
very much attracted by the idea of Borchers that the S-matrix does not
distinguish a particular local interpolating field, but rather is an object
belonging to a whole class of such fields [1],[2]. It was of course easy to guess 
that Borcher's equivalence class of all fields which are relatively local with respect to
the free field in Fock-space was just given by all local Wick polynomials
including arbitrary space-time derivatives. Although this class belongs to
the trivial S-matrix $S=1$, it was not so easy to prove this guess. The statement
(but not the proof)\footnote{ The proof I share with H. Epstein.}
 with references later entered the book of Streater
and Wightman, one of the early accounts of general QFT [2]. 

The insensitivety
of the S-matrix against local changes of interpolating fields was understood
 in a very natural way within the Haag-Kastler net-framework. It served as the
first strong hint that the new philosophy underlying the nets of observable
von Neumann algebras with their relations and inclusions (i.e. the adaption of
Leibniz' ideas about monades and the  reality created by their relations
rather than the Newtonian viewpoint of a space-time manifold and its material
content) was on the right track. 

Related to this, one interesting episode
comes to my mind. When Haag and I mentioned these flexible properties of
interpolating fields (in connection with the axial current and its use as
a pion-field) in a private discussion with Gell-Mann, he at first seemed
to be very surprised. On the next day he already fully accepted that one
does not need a Lagrangian $\pi$-field in order to describe pions.

Much later I was fascinated by Gell-Mann's idea that currents alone may
already determine the fields. I interpreted this naively (i.e. not in the
sense of the DHR representation theory of local observables) as meaning that
it should be possible to reconstruct bilocals from composite locals.
The idea worked, if one uses light-like limiting procedures [6]. 
Recently Rehren [7]
found an analogous problem in chiral conformal QFT, but in that case he had
to invent a different method.

I came to Champaign-Urbana (University of Illinois) in 1960, after having
graduated from the University of Hamburg, in order to join Rudolf Haag
(who became full professor of physics at the University of Illinois some
month before I arrived). I only had a German diploma in theoretical physics
and my credentials nowadays would have been considered as completely
insufficient for a research associateship. It was certainly not any intellectual
brilliance on my side, nor was it solely the high scientific reputation
which Haag among collegues of his own generation was enjoying. The generous
funding which physics profited from in the US during the 60$^s$ was a result
of the cold war arms race in general and the sputnik shock in particular.

Even in this ``golden age" of theoretical
physics it was not so easy to make ones carrier in such an esoteric 
(as such it appeared to the majority of theoretical physicists) area as
``axiomatic" field theory (a name which I found completey misleading) or in
algebraic QFT, where you needed a very good command of von Neumann and
$C^*$ algebras (at that time a rather dry looking mathematical  area with
little physical intuitive appeal, this was long before Connes, Haagerup, Jones, 
and others revolutionized that area). Nevertheless the contributions of Araki [1]
in these early days were quite impressive, albeit out of reach for me. 

At that time (after the aborted S-matrix bootstrap approach to elementary
particle physics) there was a strong revival of Lagrangian field theory
and renormalized perturbation theory on a more sophisticated level than that of the
early 50$^s$. At one of the meetings on general QFT, I remember Res Jost saying:
``Auf Wiedersehen in der Herberge zur Lagrangeschen Feldtheorie". Many years
before, he had said ``In the thirties, under the demoralizing influence of
quantum-theoretic perturbation theory, the mathematics required of a 
theoretical physicist was reduced to a rudimentary knowledge of the Latin
and Greek alphabets." I like somewhat provocative statements, especially if
they are able to condense the change of Zeitgeist in a perfect way. 

Somewhere
between Lagrangian and algebraic QFT there was so called ``constructive QFT",[8]
which at times tended to be also somewhat destructive.
Constructive QFT was too much limited by the idea that one only needed
a mathematical control of already more or less existing structures.
In other words, one only had to grab into that ``unclean" (from the
point of view of quantum fields and their short-distance properties)
Lagrangian box
and mathematically polish some representative models. 

It often failed to see
new structure (solitons, integrability etc.), but in many cases it
suceeded to incorporate those new things.
 
And, last but not least,
there was the newly developing area of exactly solvable two-dimensional
relativistic QFT's of which the first one was the massless Thirring model
(and the closely related Luttinger model in condensed matter physics).
I moved freely between these areas with a slight preference for two-dimensional
models and Lagrangian renormalization theory. 

The two-dimensional theories 
were called either ``trivial" or ``pathological" by the majority of self-styled
``real" physicists. But in order to make a certain specific, conceptually
interesting point (capable of a generalization to mathematically less
controllable situations like the infrared problem in QED), I  found even some of 
the ``trivial" ones useful. In this way I came across the interesting
looking concept of ``infraparticles" [9] which suggested a new scattering theory
outside the LSZ S-matrix framework. 

I never admitted the existence of
``pathological" two-dimensional QFT (at least if they, unlike the so called
generalized free fields, admitted all the known local field theoretic
structures with the possible exception of bosonic or fermionic space-like
commutation relations). It is now clear to everybody, that what some people
called ``pathological"  at that time was nothing else but : ``as free as possible
under the constraint of possessing ``exotic" space-like commutation relations".
In modern parlance they were ``anyons" i.e. fields with abelian braid-group
commutation relations and relative bosonic relations with respect to the 
observable subalgebra generated by them. 

The first paper in which an additional
parameter in the massless Thirring model was discovered which turned out to be
directly relatable to exotic  statistics, was that of Klaiber [10]. He showed also
that the anyonic statistic parameter determines the behaviour under 
Lorentz-transformations,
\footnote{A very good account about ``generalized statistics"
up to 1980 may be found in Swieca's 1980 XVII Karpacz winter school lectures.}
i.e. a new kind of spin-statistics connection valid for that extended 
Thirring model. In addition one finds half of the ``bosonization" in terms
of line integrals over chiral currents. Much later this was  independently 
discovered in a more geometrical gauge theoretical setting by Mandelstam, with a significant 
extension by Coleman. 
This interesting development had its counterpart for the one-dimensional
electron gas in condensed matter physics (the Luttinger model). The
hamiltonian bosonization (anticipating the Sugawara form of the energy-momentum
tensor for conformal current algebras) was found by Mattis and Lieb before
Klaiber's paper and the bosonic representation of fermions is due to Luther
and Peschel. Only after 1964 the two communities took notice of each other.

Lehmann [11] believed that Coleman's somewhat formal arguments
(concerning the infrared aspects of perturbations on massless situations)
are only valid in the regime of small Sine-Gordon couplings, and that the 
bosonization formulas change significantly in the larger coupling regimes.
He, together with Stehr, exemplified this suspicion by directly bosonizing  the free massive
Dirac theory without the use of Coleman's arguments. In collaboration with
Truong [12], I succeeded to understand the mechanism behind this modification of
the Klaiber-Mandelstam-Coleman bosonization as a result of the appearance
of nonleading short-distance singularities in the larger coupling regimes:
``cumulative mass effects". Later, after the renaissance of 2-d conformal
field theory and Zamolodchikov's successful 
idea to construct integrable representatives
(within the family of field theories obtained by perturbing conformal field
theories with relevant operators), it became clear to me that those
observations on cumulative effects really mean that the underlying field
algebras change dramatically under such perturbations: a (composite) primary
conformal field is accompanied by a whole family of ``shadow operators" [13]
(not conformal secondaries!),
which have no place and no name (and do not occur) in the conformal theory. So the naive
picture, that perturbation just means some sort of ``dressing" 
via states only (and that it leaves the pure algebraic structures 
unchanged) does not hold.

Jumping back to the late 60$^s$, it is worthwhile to recall, that at that time
there were many purely theoretically motivated ideas around. There was also
a potential richness of mathematical methods since the formalism of QFT was not yet
narrowed down to functional integrals (as it has been nowdays by looking at the
majority of QFT textbooks). 

One problem which attracted my interest was
that posed by Wightman [14]: to get a good understanding of stationary and time 
dependent external potential problems for free fields, including higher spin
field equations. The algebraic part of the problem was reducible to classical
retarded (or advanced) - propagators and if the external potential was
leading to a modification of the highest derivatives, then there was a 
violation of Minkowki-space causality. As anticipated earlier by Pauli and
Fierz, later exemplified by Velo and Zwanziger and slightly generalized by us
[15], this causality ``pathology" is generic for
higher spins $s\ge 3/2$. 

Many years later, Wess told me that he expects
that the restriction to supersymmetric external field problems may yield 
a higher spin situation which is consistent with causality as well as
probability conservation (unitarity). I never tried to verify whether such 
an exception to our generic findings really was possible. But is seems to
me that the failure of supergravity to lead to full renormalizability 
(of all orders, not just a high-energy improvement in lowest order) would
cast doubt on such a conjecture. 

As the important part of our external
field work we considered the understanding of sufficient conditions 
(sometimes also necessary and sufficient) for the existence of unitary
time development operators and the S-matrix by explicit formulas. Although
in our formulation the infinite Feynman phase (the one coming from the
``vacuum bubble diagram", say in a fermion theory) did not appear, it made 
its reappearance as a finite phase after concatenation of two time dependent
processes. These cocycle phases reappear in the later work of G. Segal who
uses a geometrically much more appealing mathematical framework (infinite
Grassmannians, a more geometric understanding of ``filling the Dirac-sea").

But in physics one is forced by the quantum principles to at least start
with $C^*$-algebra concepts and states on those algebras. From a mathematical 
physics point of view this problem was layed to rest by Ruijsenaars [17]
and Fredenhagen [18]. It would be a grave omission not to quote the profound
earlier work by Araki on the representation theory of CAR algebras. In a very
recent paper of B\"ockenhauer [18], this formalism was successfully used in order to
give explicit constructions of the endomorphisms (local and global) of the 
conformal chiral Ising field theory (and to show various equivalences between these 
endomorphisms). 

The extension of the Wigner theory to interacting problems with either
external or quantized fields had to be handled with care [15].
In most of the papers which followed Wightman's external field program this
care was observed, but there are also older papers where the central
issue was the covariance of Feynman rules for the S-matrix and problems
of causality and stability were not yet considered [15]. 

There was another useful nonperturbative mathematical
technique: that of spectral representations (K\"allen-Lehmann, Jost-Lehmann-Dyson).
Swieca used it in a fascinating way, obtaining a non-Lagrangian proof [19] of
the Goldstone-Theorem (far away from quasiclassical ways of thinking).
Much later he used such spectral techniques in order to prove his 
``free charge versus screening" theorem [20], a structural theorem on abelian
gauge theories. This extremely seminal paper gave rise to sophisticated 
studies within algebraic QFT of the relation between mass gaps and the best
possible localization of states, and in this way it led to what is now called the
Buchholz-Fredenhagen theory [1] (it generalizes the localization aspects of
the DHR theory). For a nonexpert it is not misleading to think of 
charge-carrying objects with the BF-localization as some semi-infinite
Mandelstam-string of local gauge theory. However, all attempts failed to obtain
an intrinsic i.e. structural insight into what may be behind nonabelian
Lagrangian gauge theories.

One should add that techniques of spectral representations were widely used
in the 70$^s$ in problems of current algebras, in particular the quantum Noether
relations between current densities and symmetry generators. They of course also 
became standard in condensed matter physics.

The most impressive theoretical discovery of those times was that of the
relation between real-time Minkowski-space QFT and euclidean QFT [8]. In a more
physically restricted sense it is the relation between noncommutative but
local QFT and certain classical  (commutative) stochastic theories defined
by Boltzmann-Gibbs formulas extended to correlation functions. These two theories
were based on completely different physical principles, and yet they had this
obviously deep relation to each other. Moreover this relation was a precise
correspondence between a quantum theory and a classical theory unlike the
various quantization recipes. Whereas quantization for certain finite-degree
systems allowed a reasonable good mathematical understanding in terms of
geometrical and topological concepts (``geometric quantization"), mathematicians
did not find a good ``drawer" in which they could place this new correspondence.
Some formal aspects were already known to Schwinger, before Symanzik, Nelson, 
Guerra, Osterwalder and Schrader (and some others) made their important
contributions. 

Schwinger had a very difficult formalism which, in contrast
to Feynman's simple rules was not overtly perturbative. It could for
example handle problems like $\mu\bar\mu$ pair production in strong fields
(of astrophysical relevance) and unlike  the aforementioned external field
treatment you did not have to break your had about how to formulate adiabatic
boundary conditions (switching on and off external fields): they were
already built into the formalism. I never completely understood all that
magic, but his former students like Kadanoff, Martin, Baym and Summerfield
were very good at handling it. 

Schwinger was the first physicist who realized that the {\bf Z}$_2$ of fermions
(showing up in the transformation properties as well as in the space-like
(anti) commutation relations) can be encoded into the position of euclidean
operators (i.e. Grassmann algebras instead of analytic monodromies in 
analytically continued correlation functions). 

Much later, Marino and Swieca [21] (in their functional
integral treatment of order-disorder variables and generalized statistics) 
tried to generalize this to the e.g. {\bf Z}$_3$ structure of $s={1\over 3}$
which occurs in 2-d models. In order to encode this into pure algebraic terms,
one would have to invent multivalued algebras, and no standard linear way of
writing products (neither Latin nor Hebrew) would allow such an exotic
structure (one would have to write into several directions simultaneously).

I was reminded of these interesting discussions I had with Swieca, when I
attended a recent seminar talk of R. Kerner with the title ``Ternary
structures and new models of gauge theories".

It is hard to imagine that Schwinger, who was a quantum realist, would have
been interested in ``Berezin integration" or other attempts of geometric
quantization, in which one invents a classical reality just for the benefit
of being able to say: I quantized something. However I can imagine that the
exploration of perturbative $\varepsilon$-neighbourhoods of free bosons,
fermions and plektons (explained later) could have attracted his interest,
especially if (as it happens in some cases) a global control is possible.

That part, which had to do with  the more profound aspects of
``euclideanization",
however, became only clear in the work of the previously mentioned authors. Under
fortunate circumstances (proximity to a $\phi^4$ model theory), the corresponding
euclidean theory admitted a Feynman-Kac functional integral representation.
The other way around the problem is more difficult. With other words, you
could not simply define a noncommutative QFT by such stochastic Feynman-Kac
type integrals, without going through a check-list of rather difficult 
properties. Only after you have done this, you are assured of a quantum
theory. The mere existence of the infinite-dimensional stochastic integrals
is not enough (and in case of nonexistence, one may still have ``structural
correctness" of the F.K. representation after renormalization, as it happens
for $\phi^4$ in $d=3$ or 4 space-time dimensions). 
The separation of  structurally correct functional integrals into algebras
(the ``Faraday-Maxwell-Einstein" local part) and states (as a result of
coherences and correlations necessarily nonlocal) belongs to one of the  most
subtle procedures of QFT. We will return to this
interesting issue in connection with ``topological QFT" via Chern-Simons
actions in the next section.

In this context it may be interesting to make the following remarks.
In the process of conversion of functional integrals and their geometric
structures into quantum physics, the BRST formalism is supposed to play an
important role. Whereas this is certainly the case in perturbation theory
(this is the origin of this formalism), one can have serious doubts about
its widespread non-perturbative geometric use (these doubts I share with Stora,
at least when I met him the last time some years ago). ``Free" fields
(including integrable and conformal fields) do not care much about the 
Hilbert-space
structure,
\footnote{Integrable  and conformal models can be mathematically controlled by only using
locality.}
they are also compatible with indefinite metric. However, it is
very improbable that interacting theories with indefinite metric can be
controlled by such methods. I do not know any trustworthy mathematical
framework which allows us to do that. The use of the Gupta-Bleuler
formalism for recovering positivity is perturbative-inductive. 

Having said that I should not hide the fact that I once commited a sin against
these principle of never touching gauge dependent objects outside integrable
(solvable) models [23]. There was the issue of ``screening versus confinement"
which was important even in 2-d models [24]. The massless Schwinger model exhibits
screening since a gauge invariant version of ``quark " fields existed
(a Dirac-spinor with a semi-infinite Mandelstam string), but it was neutral.
It was clear that the massive version (which is not solvable) behaves significantly
different. The most radical expectation was that such confining
quark operators seize to
exist as mathematical objects, since there is no physical role for them.
We (in collaboration with K. Rothe) succeeded to show that the long distance
behaviour of quark fields in the Lorentz gauge is indeed terrible, the correlation
functions are non-tempered for long distances. This was done by a tricky use
of special nonperturbative methods and, in agreement with the previous remarks, there
was no chance to use BRST for this non-integrable model. I still think that
there is some intrinsic message behind this, but  I have not been able to
liberate it from the gauge stuff.
As far as ``sinning" against the above principle, I found myself in the
good company of Strocchi and Wightman and also of Fr\"ohlich.

The real-time-versus-euclidean relation was certainly the cleanest and  most
useful result coming from constructive QFT. It was primarily a structural
insight which did not immediatly facilitate an analytic understanding of
nonperturbative QFT. However, it gave an excellent framework for the formulation
of renormalization group ideas either \`a la Wilson and Kadanoff or in the more
field theoretic spirit of parametric differential equation \`a la 
Callan-Symanzik and Gell-Mann-Low. With the help of ordinary renormalized
perturbative QFT and a trick of analytic continutation in the space-time
dimension (the $\varepsilon$-expansion with $d=4-\varepsilon$)
one could (even without knowing very much about the physics of critical phenomena)
produce some surprisingly good numbers.
Even though I contributed one of the first papers [22] after Wilson, I still
do not understand why these methods deserved such good numerical results.

Already in the early 70$^s$ there was some hope that if one combines the
acquired knowledge about the real-time-versus-euclidean connection together with
conformal invariance, one may be able to obtain explicit nonperturbative
solutions or at least to find  some detailed classification scheme of such
theories. These ideas originated from two observations. One observation
was that two-dimensional statistical mechanic models (as the Ising model
solved by Onsager and Kaufman) at criticality showed  not only scale-invariance,
but even full M\"obius-invariance, at least to the degree to which one
could understand such concepts on a lattice. From continuous  renormalized
perturbation theory one already knew that the additional conservation law
which results from being at a Gell-Mann-Low fixed point does not only imply
scale invariance but also full conformal invariance. Based on these
observations, Migdal and Polyakov suggested a ``bootstrap program" for conformally
invariant theories.  Mack and Symanzik [25]  showed that such a program basically
amounts  to nonlinear Schwinger-Dyson equations with conformally invariant
boundary condition. Mack [26] developed a group theoretical technique of euclidean
conformal partial wave decomposition, but the nonlinear ``dynamite"  remained 
resistant against an analytic solution for another ten years. 

The second
observation was that there are nontrivial massless solvable real-time
models like the Thirring model. Their observable currents are conformally covariant 
in the standard sense, but  there was an apparent obstruction against
global conformal invariance in the charge carrying fields. This obstruction was easily
related to a breakdown of Huygens principle, i.e. a kind of time-like
``reverberation", showing up in the fermionic time-like commutators.
Since this contradicted our [27] naive expectation about global conformal transformations
leading from space-time via light-like infinity to time-like,
and since it was not visible in the infinitesimal
behaviour, we called it the ``Einstein causality paradoxon" of globally conformal
invariant QFT's. We were convinced that this obstruction is worthwhile to
resolve, since it was not there in the euclidean approach and since in
our view the real-time-euclidean connection was very  deep and , as especially
the Nelson-Symanzik work showed, not fully described by only thinking
about analytic continuation. So we expected to get an additional structural
property which remained hidden to the ``euclidean eye". 

Indeed, when
Swieca  and I finally solved this paradoxon two years later [28], we had an extremely
rich harvest: a conformal decomposition theory of local fields which, when
specialized to two dimensions, is equivalent to the block-decompositon theory of
Belavin, Polyakov and Zamolodchikov. The local fields which are
irreducible under infinitesimal M\"obius-transformations, turn out to be
reducible with respect to the center of the covering group for $d=2$,
and the center
(as well as the whole group and the  theory) factorizes into the two
chiral parts. BPZ, who worked in the euclidean framework, had to obtain
this additional piece of information (which resulted in our case from the
resolution of the paradox) by the explicit use of the representation theory
of the Virasoro algebra. 

Either way, this was the ``magic" by which the
``nonlinear dynamite" of Schwinger-Dyson structures became linearized in
two-dimensional conformal field theories. 

Our findings in 74/75 confronted
us with a very perplexing situation. We thought that the Wightman theory
was a universal framework of local fields, but these new non-local-looking
fields, which came from the decomposition of the local ones with the
help of central projectors (between which the local field get ``sandwiched")
were not strictly speaking Wightman fields. They violated the Reeh-Schlieder
theorem since they had huge null-spaces (i.e. they are only nonvanishing
if applied to appropriate superselection-sectors). 

As already mentioned,
with a rich and partially-known representative theory of the Virasoro algebra
at their hand, and with the significant insight of Kadanoff into the structure
of critical points by the ``Coulomb-representation", BPZ in 1984 obtained the
first really nontrivial (non-abelian fusion laws) family of ``minimal"
models. We in 1974, lacking such powerful analytic tools (which allows one to
bypass or postpone  a lot of conceptual problems), had only the well-known
abelian family of exponential bose fields (known from bosonization) at our
disposal [28] which are too poor to serve as good illustrations of the richness of
our conformal decomposition framework. 

In the second half of the 80$^s$
when Rehren and I looked back at the old things in order to understand
their relation to the BPZ discoveries, we convinced ourselves that by
classifying statistics (i.e. R-matrix commutation relations) or fusion laws,
and computing from that dimensional trajectories of composite fields, one
can obtain a situation with a unique solution of a Riemann-monodromy problem
for the 4-point function by not using those BPZ tools but rather standard
methods of QFT. This shows that in principle we could have done it in
74, but, ``like Gorbatchev  used to say" -- who comes too early 
(and is stained by ``axiomatics") is punished by life.

Because of these conceptual difficulties posed by those exotic looking
fields and as a result of a strong prejudice in favour of euclidean
Feynman-Kac representations (I am nowadays free of that) we gave up in 75
and turned to other problems which were more in the vein of the times.
But it is worthwhile to add two remarks. In January 1974 at the V Brazilian
Symposium on Theoretical Physics, I presented a theorem [29] on the ``Lie-field"
structure of the chiral conformal energy momentum tensor. I was totally
unaware  of previous work of Virasoro who found  the same algebraic structure
by solving a constraint problem which appeared in Veneziano's dual model.
His algebra served as a kind of Gupta-Bleuler condition removing unphysical
degrees of freedom. I never used a nonlocal Fourier-decomposition. Its use
was more natural from the compact picture description of string theorist,
but a bit unnatural from the point of view of critical statistical mechanics
or  QFT. 

My motivation came from Lowenstein's previous attempts to classify
``Lie-fields" by their space-time commutation relations. Lowenstein's
 program was  probably
a precursor of the W-algebra program. In any case, the conformal energy
momentum tensor was, if I remember correctly, the first really interesting
and explicit example of a Lie-field in the sense of Lowenstein. In the same publication I also 
studied differential identities coming from short-distance expansions of chiral
currents with fields. Such differential identities are special cases of
what nowadays is called Knizhnik-Zamolodchikov equation. I used them for
the rather modest aim to show that the new solution (at that time) of the
``generalized $U(n)$ massless Thirring model" proposed by Dashen and Frishman
[30]
was not really new, since one could obtain it by interchanging the role of
Euler-Lagrange equations (which have a classical limit) with those 
differential identities (which are of pure quantum origin)
 but just starting from the old
solution. 

Those conformal models which are nowadays  called WZWN-models
in those days were called generalized Thirring models with their affiliated
(nonabelian) current algebras. 

Whereas I can reconcile myself easily with the
terminology in the first mentioned case, because Knishnik and Zamolodchikov
significantly enriched the old physical structures and produced new results, I have
some problems with the WZWN terminology in physics
(but not with the authors!). The reason is that neither is there
a close connection between those Lagrangians in lower dimensions which have
the same (or in this case a similar) algebraic form as the higher dimensional
phenomenologically motivated
WZ models (the vast difference between  the Schwinger model and 4-d QED
may serve as an example), nor can such nice looking geometric Lagrangians
through their Feynman-Kac integrals be dealt with in any convincing and
efficient way, even if these integrals would exist as mathematical entities.
There is no such problem with the old terminology, especially if one treats
such generalized Thirring models and their associated current algebras with
the help of Knizhnik-Zamolodchikov differential identities (the classical
Euler-Lagrange equations are not enough for an efficient analytical control).
Already in the late 70$^s$ it was clear that the non-abelian Schwinger-determinant
(the fermionic determinant in a generic nonabelian gauge field) was an
interesting object. We computed it in the Pauli-representation [31]. Using a
particular parametrization for the gauge potential (the Poliakov-Wiegmann
representation of 2-d gauge configurations in terms of sigma-model variables),
the Pauli-parameter is turned into a ``third dimension" and thus the Schwinger
determinant becomes local [24]. Its logarithm was some years later called the 
WZWN action. The natural development would have been the use of that determinant
for the nonabelian version of Coleman's bosonization of the U(n) 
Thirring model (the general Thirring model with arbitrary multicomponent
quadrilinear and bilinear terms is one of the richest models for $d=2$
phase transition with many conformally invariant points in parameter space).
After all the indications that this \footnote{using e.g. the quadratic completion functional
trick to convert the generalized Thirring model into a trilinear gauge
interaction, just as one does the functional bosonization of the abelian
Thirring model with the help of the usual Schwinger determinant}
may lead to interesting conformally invariant points (or submanifolds
in the general Thirring coupling constant space) already existed [30].
But the more quantum physical ideas of the seventies got disconnected from
the WZWN geometric quantization approach [32] of the 80$^s$.
Actually Witten connected the model with conformal current algebras, which
can be interpreted as being half-way towards a nonabelian Coleman bosonization
of the nonabelian Thirring model. A complete treatment may also reveal that
anomalies and obstructions are not intrinsic properties, since the generalized
Thirring model does not seem to have them.

In conformal QFT the use of the new terminology is more ideological
(in the spirit the first footnote) than computational (everybody uses old
fashioned current algebras for computations). The terminology sacrificed the
relations to the generalized Thirring model, one of the physically most
important class of two-dimensional models, in favour of geometrical prowess.
This is not the only time that physical insight was traded for ``geometrical beauty".

Algebraic QFT is one of the few areas where an impressive caution is observed
if it comes to terminology. In case of new concepts, the terminology
(at least in all cases I am aware of) is extremely clear and physically appropriate.
Take for example the notion of ``statistical dimension" in the DHR theory.
It immediatly reveals its meaning as an amplification (or multiplicity)-factor
resulting from particle (or field)-statistics. It is precisely in this way
that an experimentalizer measures the effects of symmetry (amplification factors
in cross-sections) even before a theoretician tells him how to interpret it.
Compare it with the word ``quantum group" or ``quantum dimension". Here (apart
from the misleading word ``group") one does not know whether ``quantum" is
meant in the sense of Planck and Heisenberg (discreteness) or in the sense of
Flato, Lichnerowicz and Sternheimer (deforming Poisson-brackets to brackets
of noncommutative operator algebras). The latter meaning is not intrinsic,
\footnote{If we got used to such weird terminology as ``an applied mathematician"
(or physicist!), we might as well accept the above terminological ambiguities.}
 one
will never see a quantum theory running around with a tag ``I come from
geometric quantization (or $q$-deformation)". 

This precision and depth as well as an
awareness of history one also finds in pure mathematics when it has not been 
influenced by fashion from physics. The name ``Markov trace" is an excellent
illustration. It combines the Markov of this century (the topologist) with
his father (the probabilist). 
Compare this with the above re-naming of the nonabelian Schwinger determinant.

Times of crisis usually reveal themselves
 (ever since the famous tower of Babel) through a sloppiness in
terminology.

Apart from its intermediate role in functional tricks of nonabelian
bosonization, QCD$_2$ is a fascinating
quantum physical theoretical laboratory in its own right. It was never
solved as its abelian counterpart was, but by using (uncontrollable) approximations
it was used for the study of possible confinement-mechanisms. In my view, however, 
it should serve for the opposite mechanism: the liberation of Coleman
half-space kinks. I was unable to understand such a mechanism in QCD$_2$,
but in the less interesting case of a generalized Schwinger model (still abelian)
an argument was found [33]. 

When Fredenhagen recently elaborated a general algebraic
framework for halfspace kinks [34], I was reminded of those old attempts. Since the
Fredenhagen theory is the one-dimensional version of the Buchholz-Fredenhagen
semiinfinite strings (or rather space-like cones), which everybody 
intuitively interpretes
as the quantum version of the Mandelstam strings of semiclassical gauge theory,
my conviction that QCD$_2$ is the theory of liberated Coleman-Fredenhagen
half-space kinks was strengthened. With the help of some formal progress 
coming from conformal QFT, one should be able to understand this problem.
The progress could result from using (in addition to the Euler-Lagrange equation)
those previously mentioned differential identities but now in their gauge
covariant version, which leads to a much richer algebra. I think that if 
somebody revisits those old problems with the hindsight of these new ideas,
he may be rewarded by a rich physical harvest.

It may be interesting to add the following remarks. The properties of
Coleman-Fredenhagen half-line kinks are closely related to particles which
are statistical ``schizons" [35]. The simplest example of this phenomenon is provided
by the two-dimensional massive free Dirac theory. In its Hilbert space one 
finds a bosonic field which, although beeing very nonlocal with respect to the
Dirac field (viz. the  kink-like commutation relations) still is local
relative to the observables generated by the current operator. This field
interpolates the same Wigner-particle as the Dirac spinor, i.e. the particle
is a statistical schizon. This phenomenon is related to the fact that the
algebra generated by the current does not fulfill Haag duality
([1], III 4.2, where the free Dirac current is discussed). The various
possibilities for Haag-dual extensions of this algebra give a powerful tool for
a future more profound understanding of this ``schizon" phenomenon. Note
that it is not to be confused with the bosonization-fermionization formalism
in massless theories. In the latter case the statistics of fields is fixed
by the superselection structure and one only changes the description and not
the content.

In 1976 I spent a short time in Hamburg. L\"uscher and Mack, knowing my work
and my interest in ``Lie fields" and the conformal energy-momentum tensor,
showed me some incomplete results on the representation theory of the algebra
generated by the energy momentum tensor. They had the beginnings of the
$c$-quantization for $c<1$ in their hands. I do not remember how far they
went beyond the Ising value toward $c=1$. Their paper, which went unpublished
is sometimes quoted for contributing the structure theorem of the enery-momentum
tensor. 
But (to the degree that the quotation of not generally accessible unpublished
work is helpful at all), it should be quoted as a precursor of the 
Friedan-Qiu-Shenker-work of 1985 because that was precisely what their
new contribution consisted in. Only much later, after I read the FQS paper,
I could fully appreciate what L\"uscher and Mack in 1976 had in their hands.
The FQS paper is the perfect example of deep analytic work (using the mathematics
of V. Kac) and superb knowledge of the physics of critical 2-d models blended
with a very good use of computers. However, knowing the end of the story,
one is also able to admire its ``dawn" in the unfortunately unpublished LM work.

Apart from these very special (and perhaps even somewhat premature) algebraic
contributions, the decade of the 70$^s$, as far as mathematical physics was
concerned, was definitely that of  the functional integral and the geometric
and topological structures which it suggests. I fully understoond the enthusiasm
about those new things,
since I shared it. My reasonably good knowledge of two-dimensional models,
especially about the Schwinger-model through the  1971 paper of
Lowenstein and Swieca [36] (who used operator-algebra methods) gave me an easy 
start. I was obsessed by the idea to redo everything within the new setting of
euclidean functional integrals, paying utmost attention to winding numbers etc.
In collaboration with N.K. Nielsen [37] and with some suggestions from Swieca,
everything worked perfectly (a little bit too perfect!) : one could explicitly
solve the zero mode euclidean Dirac equation in a generic abelian  gauge configuration,
see the relation between euclidean spinorial zero modes and winding numbers
explicitly in front of ones eye, derive the effective action as a conspiracy between
these zero modes and corresponding expressions in the modified fermionic
determinants, and in this way cast a new (more geometric) light on the
origin of the Lowenstein-Swieca $\theta$ angle and their description of the
spontaneous (in the sense of Higgs-Schwinger) massive breaking of chiral
symmetry and the connection with cluster properties 
(irreducibility of the $\theta$-vacuum).  The  Schwinger model was (apart
from some slight generalization) the only model for which you could do such
things (as solving Dirac equations in generic gauge field configurations).
 Since the methods of constructive field theory for calculating fermion
determinates are incompatible with the presence of winding numbers, we used 
of course the $\xi$-function method. We were attacked by Patrasciou and Seiler. 
Inspite of the fact that in those models everything was explicit and therefore
you could see in front of your eye that no principle was beeing violated,
we painfully wrote up our arguments in form of a short publication
and we thought that a title like [38]``Still more about..." would help to
close the issue (it did not). 

At that time (75/76) I was totally ignorant about the Atiyah-Singer index
theorem, although I had the conviction that there was something general behind,
maybe even already known to mathematicians. The CERN library was not good
for mathematics, so after some time I went to the mathematics library at
Geneva University where I eventually found the articles by Atiyah and Singer
and I copied some. It took me some more month in order to obtain a first
incomplete understanding. The help of R\"omer [39] (with whom I also discussed
some applications to gravity) was essential  in order to convert this
into a working knowledge. Before I knew all these things and with only my
(and Nielsen's) model calculation at hand, I gave a talk at a small meeting
in Copenhagen with some speculative remarks at the end for which I did not
have a proof, (not even good arguments) only to be reprimanded by Jackiw.

Probably other physicists had similar ideas, but in those early days 
nobody had a overall view
of what was going on. 

As far as  model applications of functional
integrals and zero modes are concerned, the book by the Abdallas and Rothe [24]
may be a good source for information. This does of course not cover the 
later more mathematically motivated papers of Witten, A. Schwartz, 
Zumino, Stora, Alvarez-Gaum\'e
and many others, in which a profound topological understanding of anomalies
was the central issue. 

Although I still followed these (especially geometrically)
beautiful developments, I did not actively participate and this was not only
because some of these things were beyond my intellectual horizon. Rather I
felt that there was an increasing gap between geometry and real quantum 
physics. I remember talks by Hirzebruch in which he explained the relations between
mathematics and ``what physicist do" (he did not say ``physics"), but I am
not sure whether he was just referring to the lack of mathematical precision
at that time (which is of course a less important issue in theoretical
physics where conceptual clarity is the important goal) or if he really
knew that physics is not always identical with what physicists are doing.
 The relevant question
would have been of course: how come that such a deep mathematical theory as
the Atiyah-Singer index theory has such an indirect relation to physical
observables? 

The index theory was applicable to euclidean functional integrals,
but e.g. the relation of zero modes to the physical issue of 
spontaneous symmetry breaking, the
Higgs-Schwinger mechanism, the condensation of chirality and disorder fields
as  well as the restauration of cluster properties etc.  was too indirect. In addition the
quantization of the classical (fibre bundles etc.) non-abelian gauge theory
was seriously impeded structurally by (even forgetting problems of existence
etc.) the Gribov obstruction. 

Now, with the hindsight of more recent 
developments, I think I have a partial answer to this question of why the
relation of quantum physics to geometry is so indirect. The reason
is that most of our quantum intuition comes from theories which allow a
``first quantization" with all its classical and geometric aspects. The
Heisenberg-Weyl algebra (Bosons) and even the CAR-algebra (Fermions),
especially with its ``reading back" into classical physics via Grassmann variables,
certainly allow for a ``geometrization". But the direct application of 
geometrical ideas may be limited to this class of theories
(this limitation seems to be shared by functional integrals i.e. the
question of Feynman-Kac representability). This is at least
the impression which one gets from algebraic QFT if one combines the ideas of
Haag, Kastler and Borchers and the DHR work with the more recent Jones index
theory. Physically such issues as states and their decomposition theory,
charges and their fusion and decompositions and statistics of particles
(in condensed matter physics also: universality classes of quasiparticles) are
more important then geometric or topologic assumptions.
 This is not to say that useful
geometric structures may not result from quantum physical principles.
But physical principles always have priority [40].
I will return to these questions in the next section.

One important development, closely related to the remarks before, ought to be
mentioned here. Already before the time when winding numbers and A-S zero
modes entered physics, it was clear that in case of breakdown of the conservation
 of the observable chiral current via a Higgs-Schwinger mechanism, the
raison d'\^etre  for the appearance of Nambu-Goldstone boson was lost and
the natural generic situation was to find a massive meson (carrying similar
quantum  numbers) instead. In fact the perturbative Higgs description
allowed to call this a ``fattened Nambu-Goldstone boson" (with Schwinger's 
composite mechanism this terminology is less appropriate). Crewther [41] tried
to use the euclidean zero modes in order to make these arguments in favour
of an $\eta$-meson more quantitative. He could not quite confirm the
expectation and he built a kind of counter-philosophy based on his 
calculations, only to get into a controversy with Coleman and t'Hooft.
The structural results of real time QFT like the H-S mechanism were simply
more trustworthy than results obtained by euclidean functional integrals
with the help of the A-S index theory. I found this a very interesting
and profitable controversy.

Since string theory is conspicously absent in these notes, I owe an explanation
to the reader why I consider it as part of a mathematical rather than a 
physical-conceptual enrichment. In fact the dual model and the old string
theory had strong physical roots: they constitute the first proposal as to what
a non-perturbative strong-interaction S-matrix could look like: there are
infinite particle-towers which in order $g\ge 2$ are turned into an infinite
tower of resonances (poles in the $2^{nd}$ Riemann  sheet).
According to my best knowledge, there is no principle in local QFT which could
forbid such a situation (i.e. there seem to be no intrinsic ``stringiness" of such a
situation). But then came the violent end: the Bartholomew-night-massacre of the old
string theory by a semantic trick. The new string theory, after many years,
got a nice geometric wrapping, but to me it lost its physical credibility.
Also I  was disappointed when the attempts about string field theory were never
directed towards the important problem of what really happens with Einstein
causality, which is the soul of standard relativistic local quantum physics.

The pitfalls I ran into with supersymmetry were of an entirely different kind.
The observations which Witten made on supersymmetrically degenerate vacua and
Nicolai's functional map (whose degree was equal to the Witten index or the vacuum
degeneracy degree) were very interesting indeed. The map was the most unconventional
way to obtain free fields which I have ever seen. The relation of that free
field to the Heisenberg fields was extremely nonlocal. Scattering theory 
provides one with interpretable free fields, which are also extremely nonlocal
with respect to the interacting fields. But for the Nicolai free fields I
could not find a physical interpretation. 

I also had some difficulties with
the vacuum degeneracy picture. In general QFT there is a theorem which allows
to decompose such a degenerate situation into a description with just one
vacuum for each component. One then expects an intrinsic property of each
component which reveals that it is coming from such a construction. I could not
find such a property, and I think there is none. Therefore supersymmetry
appeared to me as a mathematical marriage of fermions and bosons without a
visible physical match-maker. It is possible that I have overlooked something here.

\section{Recent Developments in Algebraic QFT}
The bulk of the theoretical development I am getting to now appeared
through the 80$^s$ and 90$^s$. But important ideas in most cases already
existed in the 70$^s$.

The most remarkable one is that of integrable field theories and their
S-matrix bootstrap construction. Already at the beginning of the 70$^s$
certain nonrelativistic classical field theories were shown to be integrable,
both in the sense of having infinitely many conservation laws, as well as that
their scattering theory was explicitly computable (i.e. integrable) by some
nonlinear generalization of Fourier-transformation (inverse scattering
method). The first indications
that certain relativistic theories have a particle spectrum which (similar to
the hydrogen atom) seemed to be susceptible to ``exact" quasiclassical
considerations came from the work of Dashen, Hasslacher,  Neveu, Jackiw,
Faddeev and many others. If this was interpretable as integrability, there
should be conservation laws preventing the creation of particles e.g. in 
the massive Thirring model. So this should show up as a structural property
in every order to on-shell perturbation. Berg, to whom I was an adviser,
showed the numerical absence of scattering in $2\to 3$ at a certain value
of rapidity of the Dirac particle in that model (he used a pocket computer).
 The corresponding analytic
calculations were only done half a year later, the combinatorics
(even in lowest nontrivial order) which led to this on-shell ``conspiracy"
was horrendous and one first had to find clever tricks. 

Faddeev with
coworkers, [42] with a lot of hindsight and previous work, found a closed exact
formula for the massive Thirring-model S-matrix at those values of the 
coupling constant where backscattering was absent. 

In collaboration with
Truong and Weisz, [43] I was able to show that in case of the Sine-Gordon equation
the trigonometric DHN quasiclassical mass formula for the breathers is a
result of the physical principles of unitarity, crossing symmetry together
with the assumed absence of pair creation. This was done without using the
Yang-Baxter equations which we did not know yet (for that simple scalar
scattering states one can write down the factorization into 2-particle elastic
scattering without that knowledge). 

Then Zamolodchikov [44] found a clever trick
to extrapolate the above mentioned S-matrix to the generic case with backward
scattering. By a very nontrivial combination of Zamolodchikov's result
(as a check) with a generalisation of our rather simple use of the bootstrap
 idea in that
mentioned scalar case, Karowski et al. [45] liberated the factorizable bootstrap
program in complete generality. By that time Faddeev obtained the same
structure (which he called ``Yang-Baxter") coming more from solvable models 
\`a la Baxter in statistical mechanics. 

After that, things really got boiling.
The Moscow group, notably the Zamolodchikov's  used that program and found
a tremendous wealth of integrable relativistic models, and there were 
innumerous fascinating discoveries made by Faddeev and his school. Witten
and Shankar showed how one can completely solve models (some of them supersymmetric)
by calculating the full S-matrix including that of charge-carrying solitonic
states. Swieca sent his student Kurak to Berlin in order to learn the 
``bootstrap"
trade (which he did in a short time, he together with the people at Berlin
made up tables where factorizable S-matrix were classified according to symmetry
properties [46]). Together with Swieca  [47] he later wrote a nice little paper showing
that the antiparticles in the $Z_N$ and the chiral SU(N) Gross-Neveu model
are really bound states of N-1 particles. This in turn fitted neatly with
Witten's idea how via infrared clouds 
\footnote{By an additional Schwinger-model coupling, these clouds
are ``eaten up" by the Higgs-Schwinger mechanism [48].}
the continuous symmetry breaking
(showing up in the quasiclassical treatment of these models) is evaded 
in spite of the presence of a mass gap in the S-matrix (the 
Kosterlitz-Thouless mechanism), one of the few instances of theoretical physics at its best. 

As a result of my collaboration with Swieca, I was very
much used to see solitons (in the sense of new physical charge-carriers in a 
theory which did not possess them from the beginning) and confinement (in the sense
that a Lagrangian discription may show {\it more} formal charges than the physical
Hilbert space has sectors, as in nonabelian gauge theories) as two opposite sides
of the same coin. So if, within the bootstrap framework of factorizable models,
you got very good examples for solitons, then perhaps there are also nice
illustrations of confinement. From Wigner's work on symmetries in quantum
theory one knows that the important difference to classical symmetries is that
quantum theory (as a result of its projective nature)
always asks for the covering group $\tilde G$ of a symmetry
group $G$ of automorphisms of observables. So the natural state of affairs
in QFT would be that  particles (at least the fundamental ones) also  carry
the central charges of $\tilde G$. One only had to look for integrable
models like the O(3) model where this did not happen and understand this
``exceptional" behaviour in terms of confinement. When I discussed this
idea with Kurak, he told me that Karowski had already made the formal 
observation that the O(3) S-matrix was related in a very simple way with the N=2
S-matrix of their SU(N) table. When Karowski returned to Berlin from a trip,
the paper [49] was already almost finished, he only had to correct some sign
related to the parity of bound states. 
It is well-known that this model also admits a gauge-theoretic Lagrangian
description, which is useful if one wants to study its quasiclassical aspects.

A short time later, Karowski [50] published
a very seminal paper with a systematic study  of bound states and fusion within
the framework of factorizable QFT, paying special attention to quantum
probability conservation (positivity of residua  of boundstate poles). This
paper explained, on a much more fundamental level than before, the origin
of those trigonometric mass-trajectory formulas (known since the DHN quasiclassical
work). 

In fact, with hindsight of more recent developments, their origin is formally the same as that for
the statistical dimensions in the DHR treatment of low-dimensional QFT or that of
the size of subfactors with Jones indices smaller than 4: the fusion laws
give rise to the Perron-Frobenius problem. This paper, via its later extension by
Kulish, Reshetikhin and Sklyanin [51], led directly into quantum groups.

Drinfeld discovered (in a mathematical sense) that the original  Yang-Baxter
equations, extended to a system of Yang-Baxter equations for all particle states
(i.e. including the fused or ``cabled" bound states), lends itself to be
interpreted as the representation theory of a kind of $q$-deformed group.
This was done in a general algebraic setting without paying attention to the
probability conservation of quantum theory. 

I am not quite sure, whether this
development was altogether healthy for theoretical physics. For a theoretical
physicist, the ability to manipulate quantum groups and $q$-deformations
(meanwhile of ``everything") without having a profound knowledge of QFT, is
a dangerous enterprise, reminiscend of that previous quotation of Jost.
 Positivity is not such an important issue in 
statistical mechanics. In the formulation of Jimbo, its Yang-Baxter and fusion
roots in the spirit of Karowski and Kulish, Reshetikin and Sklyanin 
remained more visible. Woronowicz got his ``pseudogroups" within a $C^*$ algebra
setting in a completely intrinsic and independent mathematical way.

According to my best present knowledge, the $q$-deformed group at generic
q-values was never incorporated into the quantum physics (Hilbert space, operators)
of the Sine-Gordon equation. Rather (similar to algebraic QFT) it was observed,
that certain objects (Markov-traces and their ensuing R-matrices), which one
can obtain by reconstruction from quantum groups at roots of unity (and which
do not show the pathologies of quantum group at those exceptional values), are
identical to structures which one attributes the the so-called ``reduced"
Sine-Gordon model (at quantized values of the Sine-Gordon coupling constant).
The present widespread use of $q$-deformations in ``physics" 
is not covered or supported by such observations.

In the study of superselection rules of low-dimensional QFT's, quantum
groups are only useful at roots of unity, when they give rise to Markov
traces on the ribbon braid group algebra {\bf C}$RB_\infty$. But the latter can
also be computed without quantum groups, using only physical concepts. The Markov
traces lead to unitary representations of {\bf C}$RB_\infty$, but unfortunately 
(at roots of unity) in the reconstructed version they ``loose their memory" 
about to which dual quantum
symmetry they want to belong to (at least for the time-being). The way back from
Drinfeld via Kulish et al. to Karowski's particle fusion is analogous to the
use of $q$-deformed groups in order to obtain Markov traces. I suggest to the reader
that he permits himself this journey into the past because it is physically
very profitable.

The name ``Yang Baxter" chosen by Faddeev, is of course historically completely
correct. But Yang's work on $\delta$-potential scattering theory 
(one should not completely forget Mc Guire and Brezin's et al. contributions
at the same time, but Yang's seem to be the clearest)
 was probably not known by Faddeev at the time of the bootstrap
discovery (he certainly knew Baxter's work). Karowski et al., on the other hand
refer to that Yang's work in their very first paper, but they did not know
Baxter's work at that time. There is however an essential amount of QFT in that
game, and a young newcomer should be aware that Yang's work would have remained
an isolated piece of low-dimensional potential scattering theory, and Baxter's
work would not have found the proper central position in quantum physics
without these fresh quantum field theoretical developments. 

I use the name
Yang-Baxter in connection with $\theta$-rapidity dependent S-matrices, but if
it comes to R-matrices describing braid group statistics, I prefer the name
Artin. It is true that formally the latter one is obtained from the other by
$\theta\to \pm\infty$ (after rescaling), but from a physical conceptual point
of view there is an enigmatic distinction. Statistics of particles is a long
range phenomenon (in principle one can study it on a lattice). Even though
in 2-d physics (in contradistinction to 3-d) particles cannot generically be
described in terms of statistics (i.e. statistics is not a generic attribute
of composition of 2-d charges) it is reasonable
to follow Swieca [47] and distinguish between an analytic S-matrix fulfilling
a Yang-Baxter equation and a ``physical" S-matrix where the discontinuity in 
the $\theta$ variable at $\pm\infty$ is placed at $\theta=0$, in order to comply
with the LSZ properties and the long-range nature of statistics. 

Jones is
one of the few mathematicians who makes a similar distinction; in his terminology 
there are the R-matrices which appear naturally for large families of subfactors,
and there is the program of ``Yang-Baxterization" which, for the time-being, has
not such a clear conceptual placement in the Jones subfactor theory.

At the time when Karowski and Thun [52] investigated the fusion-, spin- and statistics
structure of Gross-Neveu O(N)-model solitons, the University positions of
Weisz and later also Thun (Berg left immediatly after his PhD) expired
and the resulting existential problems led to a decay of the whole group.
Research on 2-d field theory at that time in Germany was not exactly
a carrier-building activity. A whole group of very bright physicists never
got a professorship in Germany.

I remember one, at that time very hard-looking problem which Weisz raised in 
connection with their work on formfactors. He realized that some of the
formulas which appeared in this work were identical to formulaes in some
statistical mechanic work of Lieb. Lieb appeared in Berlin quite frequently
and so at the next visit Weisz asked that question. Lieb also did not have
an explanation. This problem was finally solved by Zamolodchikov, and its solution brought
QFT and statistical mechanics still a little bit closer. 

One of the most seminal papers on two dimensional QFT's of the 80$^s$ is the
already mentioned BPZ work on conformal QFT. With the powerful tool of 
representation theory of the Virasoro algebra, this time interpreted as the
Fourier-components of the energy momentum tensor (the original interpretation
in terms of a constrain algebra was maintained only in string theory), they
calculated a fascinating new family: the minimal models. This work was
significantly extended by Capelli, Itzykson and Zuber, and further developed
and explored for the benefit of critical phenomena by Cardy. In this post - BPZ
work, the concept of modular invariance by Witten and Gepner, which came from
string theory, played an important role. 

Around 1986, I thought that it would
be worthwhile to develop this (in the spirit of the 74/75 work of Swieca
and myself) more into an algebraic direction, with the hope to ``liberate"
some of the ideas from their narrow conformal compound. In our old work,
we did not make the relation to the DHR theory of superselection sectors although
(we were both closely linked to Haag at one time or another) we knew its 
main physical motivation and certain technical aspects. The only physicist, who
had the courage to use this in spite of the low reputation (``pathological")
which low-dimension QFT's were enjoying at that time, was Fr\"ohlich [53]. It is
somewhat  ironical that we did very similar things (unfortunately sometimes
 ignoring
each other) but at different times. When he worked on algebraic aspects, we
(following the order-disorder duality ideas of Kadanoff) looked at euclidean
functional representations for sector-creating relativistic soliton 
correlation functions (we found them in the form of generalized euclidean
Aharonov-Bohm representations [54]) and when he [55] studied the euclidean approach
to solitons, I was at algebras. But on the issue of braid group commutation
relations we both met, he with the personal knowledge of Jones work and the contribution
of Kanie-Tsuchiya, and I just following the intrinsic logic of QFT (which had already
led to those 74/75 results), being ignorant about those at that time very
recent developments [56].

In those days I had a very intensive collaboration with Rehren.
We developed our framework of exchange algebras after making a very careful
check in which we had to calculate the positive definite $n$-point function of the
$d={1\over 16}$ field in the conformal Ising field theory [57]. One only had to
carry out what Kadanoff and Ceva said (mostly verbally) in their paper and
combine this with the simplicity of the ``doubled" (Dirac instead of Majorana)
model (obviously a positive Wightman theory) by drawing a kind of ``holomorphic
square root". Nevertheless the calculations meant hard work, especially for
Rehren. Only after this we had the courage  to propose the general framework
of those new algebras. This care was necessary, the fields were not 
 Wightman fields (this was already
mentioned in the previous section) and, in  relation to this, the commutation
relations were not really group-theoretic-tensorial (but rather the indices were
edges on fusion graphs) as one would have expected from a Wightman framework.
\footnote{Tensorial commutation relations are only consistent with abelian
braid group commutation relations. In some of the early work on braid
group statistics this has been overlooked.}
 The title ``Einstein causality and Artin braids" was
chosen in order to indicate the conceptual origin of these new structures
already in the title. We were able, among other things, to demonstrate that the
general principles of QFT were strong enough to obtain at least the same explicit
 insight e.g. into minimal models as the BPZ  formalism based on the
Virasoro algebra representation theory.
 We also thought that it should be possible to unravel the
rather complicated space-time structure of ``plektonic" (i.e. non-abelian
braid group representation) correlation functions by ``liberating" the Bethe
Ansatz from its statistical mechanic compound in order to get a generalized
Fock-space description, but this hope (to which I come back later in this 
section), up to now, did not materialize.

After the arrival of Fredenhagen, who joined us in Berlin (unfortunatlely only
for 3 years), it was possible within a very short time, to describe the 
exchange algebras within the general DHR framework of superselection-sectors
in such a way that finally concepts and formulas hold for all low-dimensional
QFT [58].

Parallel to this development, Witten, following the more geometric logic
(which also underlies string theory), developed the ``topological field theory". His approach
was a quantization scheme where one starts with classical actions and via the 
mediating power of functional integrals tries to obtain quantum theories.
From the classical Chern-Simons action he found, with a lot of hindsight,
the Jones polynomial or (in more technical mathematical language) the
tracial Markov state on the non-commutative infinite braid group algebra 
{\bf C}$B_\infty$ (he obtained quite a bit more, 
namely the knot invariants and braids on any Riemann surface
and, as some kind of vacuum partition functions, new invariants of general
3-manifolds). 

These findings were very surprising to us, since we encountered the 
Jones invariants in a completely different-looking context, they appeared inside
the von Neumann type $II_1$ intertwiner algebras with their Markov traces
as unitary representations of the {\bf C}$B_\infty$ braid group algebra
(the Markov traces being  the ribbon-knot  invariants analogous to the
{\bf C}$S_\infty$ invariants of the DHR methods). The invariants of 
3-manifolds which we did not have initially, were obtained later by
converting multicoloured cabled Markov traces into Kirby-invariants
[FRSII, appendix]. Here we
used the idea of some limit endomorphism $\rho_{reg}$ (or rather a limiting
tracial state for infinitely thick cables). Wenzl was able to control those
limits directly (we  only abstracted a formula by using their suggestive
power) and in this way characterize the invariants of 3-manifolds as being
uniquely related to colour-averaged Markov-states on the pure ribbon
braid group RPB$_\infty$ (the restriction to pure braids is necessary to have
a 1-1 relation). With the help of triangulation ideas, we
could also obtain the relation to the combinatorial Turaev-Viro theory
(with a precise criterion for the conditions under wich the Turaev-Viro
invariants are the absolute squares of the Witten invariant), but this looked
more artificial, at least from a physical viewpoint. 

What really surprised us is that Witten could obtain these invariants directly
in a seemingly field theoretic way, whereas if one starts from a full field
theory with localization and space-time translations one would never be able
to see a $II_1$ von Neumann algebra i.e. a tracial state on {\bf C}$RB_\infty$
directly, but only via physical Markov traces on intertwiner subalgebras
(here nothing is invented, everything is derived from physical principles!).
How can any quantum theory algebra be directly $II_1$ (i.e. an algebra
which only carries combinatorial data)? In finite-degree quantum mechanics 
one only sees such structures, if one looks at the Weyl-like algebra going
with a particle on a circle (the so-called noncommutative torus algebra,
which is not really a Weyl $C^*$-algebra in the strict mathematical sense).

In order to try to give at least a tentative answer, I have to return to some
subtleties in the issue of the Feynman-Kac integral (with an ad hoc chosen action)
defining a quantum theory and not every quantum theory being Feynman-Kac
representable. The precise conditions have been studied by Klein and Landau [59].
The quantum theory must have the property of ``stochastic positivity" (and the
stochastic theory must be Osterwalder-Schrader positive). This means that it
must admit an abelian subalgebra such that a group of automorphisms 
(physically: time translations) applied to that subalgebra generates an
algebras which is dense in the total quantum algebra. In more physical terms:
the theory must be structurally close to a canonical theory of the $\phi^4$-type
(the integrals can only be controlled in low dimensions, but for $\phi^4_4$
one has at least ``structural correctness").
\footnote{ Structural correctness cannot be checked via quasiclassical
approximations. Quasiclassical approximations cannot distinguish local
quantum physics from  ``quantized" infinite dimensional symplectic geometry.} 

This is definitely  not the case
for the Chern-Simons action. Apart from the additional factor $i$ in its
euclidean F.-K. representation, it deviates significantly in its canonical
structure e.g. in the  time-like gauge. The $\delta$-function, which one
usually encounters in the commutation relation of the field with its time-derivative
now appears between the two spatial components of the gauge field. A simpler
example of a theory which violates the Klein-Landau prerequisites, is the one
defined by just one chiral current (in this case nobody would dare to write
a F.-K. representation). 

The K.-L. theorem does of course not rule out the
possibility of a Chern-Simons quantum theory (and it does not restrict
the use of functional integrals for geometric purposes in mathematics
outside local quantum physics). So one could go ahead and try
to extract some noncommutative $C^*$-algebra via a canonical formalism and
impose some state on that algebra in order to implement gauge invariance.
This was indeed done in the abelian  case with the result that the algebra
is a kind of Weyl-algebra over 1-forms [60] and the state was that natural singular
state (thus implementing the intuitive idea of ``summing over all gauge copies")
which already appeared in earlier work. The GNS reconstruction  of the
Hilbert space representation via that singular state showed the discreteness
of type $II_1$, i.e. the space-time translation was ``killed" in the
representation obtained by that singular state, but it is still a shot apart
from Witten's theory. 

My suggestion would be to think (as an example) of the $Z_N$ abelian
Witten theory as a maximal extension of this 1-forms (as in the simpler case
of the $Z_N$ conformal field theory). But one more algebraic step has to be
carried out: the globalization of the $Z_N$ Weyl-like algebra $\cal A$ to a
universal algebra ${\cal A}_{uni}$  [58] (we will return to this in a different 
context later). It is hard to say what this means in terms of the semi-infinite
string-like 1-forms, without having done the calculations (probably it means
that in addition to exact forms  one also obtains closed forms in the angular
sense). I expect, that a suitably defined gauge invariant state on ${\cal A}_{uni}$
will give the $Z_N$ Witten's theory. 

If this were true, one would get a rather
new view on low-dimensional gauge theories. Whereas in $d=4$, gauge theory has
two aspects, the more physical ``quantum Maxwell" aspect and the singular
state aspect (``summing over $\infty$ gauge copies"), only the latter remains
in low dimensional gauge theory. 
Following Narnhofer and Thirring as well as Acerbi, Morchio and Strocchi 
[61], one should then expect the gauge invariant
algebra to be just that subalgebra on which the singular state becomes regular
(i.e. the algebraic QFT version of the geometric BRST procedure).
For topological field theories this algebra would consist of completely 
combinatorial ``stuff",  wheras if you couple the Chern-Simons gauge field
to e.g. spinor matter (extending the Weyl-like algebra in a suitable way with
a CAR algebra) then the regular  subalgebra would be expected to 
carry spatial translations
and allow localization as usual in QFT. 

So, in a novel way,  gauge invariant
singular states could generate interesting plektonic  subalgebras inside 
old-fashioned bosonic (Weyl-like) or fermionic CAR algebras. 

This then would relate to Jones theory
(Jones obtains all his subfactors within $R$, the unique hyperfinite $II_1$
factor) in a physically deep and perhaps even practically useful way by
attributing to low-dimensional gauge theories this very physical role of
finding new algebras inside old ones (i.e. something which cannot be done
``by hand"), which is the algebraic QFT counterpart of perturbing Hamiltonians.

Having said this, I should add however that I do not negate the fast and
efficient power of geometrical methods to produce new and interesting formulas
(with later chances of physical interpretation). The most famous historical
illustration is that transformation formula which Lorentz and Poincar\'e shared
with Einstein.

It is a bit regrettable that physicist have forgotten the art
to obtain new insights by pushing paradoxical-looking situations to their  
breaking point. 
The mathematical ``fuzziness" of the Feynman-Kac integrals (outside the limitations
set by ref.[59]) has propagated into a conceptual fuzziness on the physical 
side, a kind of geometrical ``dreamland".\footnote{Physical words like
particles, states, confinement, symmetry breaking, condensation etc.
do not mean what they used to mean (and still mean in this article), they
rather have become geometrical allegories of local quantum physics.}

Almost everything is covered these days by a 
confidence-creating  geometric layer. In earlier times physicists
obtained spectacular progress by confronting paradoxical situations
(viz. the Bohr atomic model) directly. When I talk these days to mathematicians
(with a few significant exceptions) they think we are sharing a happy marriage
and live jointly in a beautiful castle, they don't see that they alone live
in a castle built on our physics ruins (instantons, . , . ). 

Most of my
physics collegues agree that some parts of theoretical physics live through
a crisis, and some of them even agree that this is home-made (different from
the stagnation created by the huge phenomenological success of electro-weak
theory). 

In a recent panel discussion one could only choose between the
scylla of geometrically motivated quantization schemes and the abys of learning
how to live happily with cut off dependent real theories like QED, $\sigma$-models,
$\phi^4$ theory etc. The conceptual messages coming from the first half of
this century, that one should always aim at theories which are fulfilling all
the presently known principles (unless one incorporates them into new ones),
seems to have been lost. There is really the danger that this century ends
like this, inspite of its conceptually glorious beginning. Even the important
message (from Faraday, Maxwell and Einstein) that local structures 
(action-at-a-neighbourhood principles, locality, causality)
 are primary and global
properties have to be derived from local properties, seemed to be forgotten.
Maybe fin de siecle crises are a natural collective phenomenon. In that case
one simply would have to wait for another 10-15 years.

The main issue in that panel discussion was the funding
of (mathematical) physics, i.e. the economic crisis in physics.

In earlier days, even big physics conferences were profitable, because there
were ``art critics" who asked many penetrating questions and made 
interesting and physically relevant remarks. Pauli, Landau, K\"allen and
later also Coleman even shaped the direction of research, although sometimes
with misjudgements and prejudices.

I was moved, when Pauli, obviously seriously ill (he died 6 weeks later) after
delivering a beautiful lecture on the neutrino and its history at Hamburg
University took some rest in ``his" turning arm-chair (which we called
respectfully the ``Pauli chair"). He suffered from fatigue and pains and
he said: ``ich denke, der Heisenberg liegt mir noch schwer im Magen". He was
obviously referring to his involvement in the aborted nonlinear spinor
theory which (in contrast to the quark model) was only able to produce particle
parities which were ``correct up to a sign".
\footnote{But apart from that the underlying ideas had a certain similarity
with quark-model ideas.}

Nowadays many talks in planary sessions allow no critical feed-back from an
``art critics" physics point of view. They tend to have more similarity 
with sporting events where highly paid and often anabolically supported athlets
give their impressive performances. Since my attendance of big conferences
was rather limited (the 1994 Paris mathematical physics conference was the 
only big one I attended during the last 10 years), my impressions may however
have no statistical significance. Coleman's yearly ``physical weather reports"
from Erice still enjoy a prominent place on my shelf.

Since I do not want to end this section on such a pessimistic note, let me 
report on two interesting recent developments.

The first is contained  in a paper by Doplicher, Fredenhagen and Roberts [62] and
adresses the question whether a non-commutative version of Minkowski-space
is possible
(i.e. non-commuting position operators) which leaves the Wigner one-particle
structure essentially intact. Well, it is possible, and the physical core
of this paper is the derivation of a new uncertainty relation by confronting
the quasiclassically  interpreted Einstein equation with particle quantum
mechanics and invoking a principle of local energy stability against the evaporation
into small black holes. 

This is very appealing indeed. Fundamental uncertainty
relations are hard to get, and the last one after Heisenberg's,  was that of Bohr
and Rosenfeld for the electromagnetic fields (by consistency of quantum
particle theory with Maxwell fields). The saturation of these uncertainty
relations by a concrete model algebra is to be understood as a model
illustration. In the DFR paper, Einstein causality is replaced by something
else, not yet fully understood. Of course one wants to keep macro-causality
i.e. causality at large distance. All previous attempts to outmaneuver Einstein
causality,  e.g. those attempts to admit Lagrangians with formfactors
(on a fundamental level), or the Lee-Wick attempt to admit pairs of complex
poles in the Feynman rules, failed on macro-causality after building up increasing
orders of perturbation theory (leading to non-interpretable precursors).
The conceptual ``tighness" of Einstein causality lends credit to the belief
that any interpretable physically consistent QFT which goes beyond, must
necessarily lead towards quantum gravity. As a curious side-result of the
DFR investigation one should mention that they also  obtain 
(in their model) a two-sheeted space similar to Connes's results about the 
non-commutative geometry interpretation of electro-weak interaction phenomenology.

Another interesting proposal is the reconciliation between the renormalization
group and the resulting scaling theories with the framework of algebraic
QFT, by Buchholz and Verch [63]. I remember my difficulties I had way back, when
I wanted to understand the physical content of the Gell-Mann-Low-St\"uckelberg
renormalization group equation; one always seemed to be close to tautologies.
After Wilson's and Kadanoff's work and the Callan-Symanzik equation of QFT, this
uneasy feeling was removed, but the question remained: what is the intrinsic
physical content? The renormalization group calculations played an important
role in the construction of field theories starting e.g. from the Feynman-Kac
quantization scheme. But suppose one already has a theory, which in the
algebraic formulation would be a concrete Haag-Kastler net. Can one construct
an associated scale invariant net? Can the short-distance aspects of quarks
(i.e. their role as partons) be understood in that associated theory?
Can they be treated (in the scaling theory) as Wigner particles with perhaps
new superselection rules (``liberated color") which in the original theory
was not possible? For the answer to some of these questions we refer to the 
paper quotend above.

\section{Modest Aims, What Algebraic QFT Should be Able to Achieve in
the Next Future}

In QFT, the operator theory of free fields and their Fock-space structure
(as unsophisticated they may appear to a young generation raised with
differential geometry and algebraic topology) plays a pivotal role historically
as well as for its physical interpretation. Most physicist know free fields
through one or other form of quantization, few are aware that they are a
consequence of purely intrinsic quantum physical principles: if one starts
from Wigner's group theoretic one-particle classification and builds up
multiparticle spaces by tensoring (this being the correct formulation for
statistical independence in combining subsystems in quantum theory) and
invokes the Einstein Causality principle, one invariably comes to this
structure. It is also unavoidable if one extracts scattering theory from
Haag-Kastler nets. In the latter case one does not only find these free
fields as unbounded operators affiliated with the nets, but one also learns
that the S-matrix only depends on the nets and not on any particular interpolating
field coordinates which one may use as infinitesimal (point-like) generators
of the local algebras. 

All these are valid statements, if the theory is
four-dimensional, in which case the free fields are strictly speaking the old
``permutation-group-statistics free fields". According to the new insights
into low-dimensional QFT, this does not happen for $d<4$. There Wigner particles
may obey braid group statistics (for massive two-dimensional
theories, the generic situation would be even more ``exotic"). In this case
the powerful formalism of scattering theory leads to an asymptotic multiparticle
momentum space structure on whose unraveling considerable progress has been
made recently [64]. But in that case the causality principle (more precisely the ``localizability"
of fields, since Einstein causality only applies to observables) cannot be simply
implemented by Fourier-transformation: ``free plektons" would necessarily have a
more complicated structure then free bosons or fermions. 

Let us look at this
situation from a slightly different angle [65]. Suppose we would already know
(d arbitrary) an S-matrix with all its creation and annihilation processes.
What happens if we consider its ``extreme cluster limit" i.e. we spatially
remove all particles from each other (in such a way that no pair stays close)?
The naive answer that $S\to 1$ is the correct one in $d=4$. On the other hand
one would expect that $S_{lim}=1$ is exhausted by the Borchers class of
free fields (although this only has been proven for the case of zero mass
particles). So the cluster argument also leads to an associated free field
theory. 

Let us now try to understand such limiting theories for $d=2$.
Qualitatively one expects a similar behaviour as far as the asymptotic
supression of on shell creation and annihilation is concerned, since the
higher thresholds even in $d=2$ have more regularity properties and since smoothness in
$p$-space amounts to large-distance fall-off properties in $x$-space. However this
simplification cannot go as far as $T=0$, if $T$ denotes the excess over the
identity contribution, i.e. cluster arguments about the S-matrix in $d=2$ are not
able to separate ``interaction" from an allegedly non-interacting part
(formally: the momentum space $\delta$-functions is the same in front of
the two contributions). This is to say that it is difficult to give an
intrinsic meaning to the notion of interaction by looking at the two-particle
S-matrix (except the statement that the limiting elastic part does not show
any higher thresholds). Direct higher elastic processes like $3\to 3$ also
have stronger $x$-space fall-off properties as compared to those which happen
in subsequent stages through $2\to 2$. Hence, if the limiting S-matrix 
$S_{lim}$ is again a unitary S-matrix in its own right, then the rapidity-dependent
$S^{(2)}_{lim}$ has to fulfill the Yang-Baxter  equation as a necessary
physical consistency equation (i.e. there is nothing to be imposed from the
outside). The only surprising aspect is the claim that $S_{lim}$ again belongs 
to a localizable QFT.

For $d=3$, an educated guess would be that $S_{lim}$ inherits the energy-momentum
independence property from $d=4$, but it is only locally independent. When two
momentum- (or rather velocity-) directions cross, 
then an Artin R-matrix makes the transition from one momentum
space ``wedge" to the neighbouring one, similar to the $\theta=0$ jump 
for $d=2$ theories in the
Swieca description with $S_{phys.}$ instead of the Yang-Baxter $S_{anal.}$.
This presupposes of course that LSZ (or  Haag-Ruelle) scattering theory is
a good framework for the scattering of ``plektons". (There is a more general
formulation which uses only expectation values and not amplitudes).

The picture obtained in this way clearly leads to an identification of
``free plektons" (i.e. space-time fields belonging to the limiting theory)
and the notion of ``integrability" in QFT. It is completely consistent with
the results of Coleman and Mandula [66] which were obtained with the help of
different ideas. 

In fact it seems to be a more profound version of the
Coleman-Mandula results, since it leads to a long-distance (in the sense of
$S_{lim}$) universality-class division of low-dimensional QFT's. Such universality
classes of ``quasiparticles" (thinking of condensed matter physics where
only ``localization" is important but not Lorentz covariance of energy-momentum
dispersion laws), with precisely one ``integrable" (or free-plektonic) 
representative in each class, are extremely promising concepts with potential
experimental ramifications, at least to the extent that ``quantum layers"
are important for explaining newly observed effects. 

Algebraic QFT tells us that
the appearance of plektonic statistics and 2-d kinks is the only significant
 difference between low- and high-dimensional relativistic QFT.
This universality picture would be analogous to the short-distance universality
which, for $d=2$ led to a classification theory of critical indices in the theory
of critical behaviour. 

Most of my solid state physics collegues are not aware
of the fact that although the physics of critical phenomena is ``euclidean",
the classification method which explains the spectrum of critical indices
in terms of superselected charges, their chiral fusions rules, and their
braid group statistics (and therefore at the end,  knot-theoretical invariants)
is of course done on the noncommutative QFT side of the mysterious euclidean
statistical mechanics versus real-time QFT connection. 

Already Kadanoff, to whom
some of these concepts were not yet available (and may even have been too
exotic in order to be acceptable to him), appealed to the Coulomb-representation
although, strictly speaking superselected charges have no direct meaning
in the two-dimensional classic stochastic theory (as the value-space of
classical field manifolds has no direct meaning in QFT). 

A universality-class
theory of quasiparticles (with their spin-statistics and charge fusion properties)
in the mathematically clear form of free plektons is still a dream about the
future, but the analogy with the QFT understanding of 2d critical phenomena
gives an impression of how good life could be! 

Its origin in relativistic QFT
would by no means prevent its use in nonrelativistic condensed matter physics
inasmuch as the relativistic proof of the spin-statistics theorem and the
notion of antiparticles does not foreclose their validity in the nonrelativistic
physics of fermions and bosons. 

It is however not so easy to see whether ad hoc
introduced degrees of freedom like ``spinons" and ``holons" could naturally result
from such a more general looking scheme, which is still severely restricted
by ``localizability". The spin and statistics of nonrelativistic plektons is
more flexible than that of bosons and fermions, but there still will be a spin-statistics connection!

There is also the equally important question to what extent physical phenomena
like superconductivity (old and new) are really universal. Is it really true
that the relation between $T_c$ and the properly defined gap in a theory with
only 4-fermion interactions (the original BCS model) does not depend on the
details of the coupling (whereas $T_c$ and $\triangle$ do depend) and can only
be changed by changing the degrees of freedom of the system (e.g. by introducing
phonon degree's of freedom)? 

As a non-expert I would interpret the general acceptance
of results of approxi\-mation-methods like Hartree-Fock, random phase etc. as an
indication of universality of that relation. But then, at least in pure 
quadrilinear fermionic theories, one could only change that relation by
giving those fermions an internal, say SU(2) degree of freedom by which the
relation would get immediatly modified by a factor $2^2$. This is of cause
physical nonsense since nobody can imagine a phase transition which converts
ordinary U(1)-symmetric condensed matter into SU(2) matter. But plektons
are different, they are as selfdual as a U(1) theory (or rather $Z_N$-theory
if one quantizes the charges), even though they are more non-commutative!
In fact phase transitions away from the perturbative fermi-liquid phase
provide the only conceivable mechanism by which they can appear. 

If this is so,
then the amplification factors in the  $T_c$/gap relation should be interpretable
in terms of squares of statistical dimensions, which is the same as Jones
indices! 

In the literature one finds many geometric investigations of braid group
statistics based on the Aharonov-Bohm or the Chern-Simons theory, see for example
the work of Wilczek [67]. Since I have already commented on the difficulties in relating
Chern-Simons with quantum physics, let me now make some remarks on the use of
the Aharonov-Bohm effect. 

The physically measurable A-B phase shift of this effect
is completely independent of the subtleties one encounters if one uses this
long range A-B interaction for pair-interactions between particles. Whereas
in the first case, only quasiclassical properties are relevant, the boundary
condition which one should use in the scattering theory of such long-range interactions
are not so clear. If one wants to use the A-B theory for the description of
particle statistics and is inclined to take some hints from algebraic QFT,
then it should be chosen in such a way that the cluster property holds, i.e.
the three-particle A-B theory should contain in it the say previously understood,
two-particle theory and so on, like an infinite russian matrushka. The
correct boundary conditions should be those which are consistent with these
cluster-properties which would convert the original two-particle problem into
one with infinite degrees of freedom. 

Of course I do not know whether such
a ``bootstrap"  version of the A-B theory is feasable, the only thing I am
saying is that it ought to, if this effect
can be used to describe abelian braid group statistics i.e. ``anyons". 

I am
convinced that braid group statistics in Q.M. cannot be analyzed as an issue
separated from QFT. In the nonabelian case the above ``cluster tower" 
is reminiscent 
of the algebraic Jones tower in the subfactor theory. Using this analogy, the
cluster properties of nonabelian A-B theories should lead to a quantization similar
to Jones or to that of statistical dimension as in algebraic QFT. In QFT these
cluster or tower (tunnel) aspects are built in through the net structure, and
the localization principle for the physically admissible state on such algebras,
unlike Q.M., they do not have to be added ``by hand".

Pure geometric considerations alone, whether in the original form of Leinaas
and Myrheim, [68] or in the mathematically refined recent version by Mund 
and Schrader, [68] constitute the geometric prerequesites, but for themselves
are not sufficient in order to obtain a plektonic quantum mechanics.

The issue of ``new degrees of freedom" has also progressed significantly in
lattice models [69] of statistical mechanics. There, the ``old degrees of freedom"
which correspond to fermions and bosons of continuous QFT, are ``Paulions"
i.e. lattice spins repeated on all sites. 

In the language of quantum spin
chains one obtains
an infinite spin chain as a limit (using appropriate boundary conditions) from
finite chains. In the case of free boundary conditions this corresponds 
precisely to the constructions of a so-called UHF algebra $\lim_{n\to\infty}
Mat_2${\bf (C)}$^{\otimes n}$ from  finite full matrix  algebras, with the lattice
labels being the floors of the Bratteli tower (this only gives a semiinfinite
lattices, the full lattice can be obtained afterwards by translation). The
corresponding Bratteli diagrams for Paulions are of the most trivial type:
they have a period two, and in case of $Mat_2$({\bf C}) also 
width two. 

In a profound paper Pasquier [70] realized that (using Ocneanus ideas)
one gets a tremendous richness if one allows a width which successively
increases up to
a maximal value (related to the depth of Jones inclusions). In this way he
was able to describe the kinematical aspects of local observable algebras underlying
the so called RSOS models of Andrews, Baxter and Forrester. 

Recently this 
idea was extended by Jones (with the help of ``commuting squares", a kind
of ``pre-braid" formalism) in order to incorporate the three physically
important families into his subfactor framework: spin models, 
\footnote{ The inclusion related to the above Paulion chain is
that of Z-components of spin (diagonal matrices) 
into the full matrix algebra.} vertex (=SOS)
models and (for the new degrees of freedom) RSOS and IRF models (the latter making 
their appearance through Hadamard matrices in the commuting square scheme).
In addition, he was able to obtain a subfactor interpretation of periodic
boundary conditions which have thermodynamic limits outside the  above AFD
algebras (i.e. those which  one can obtain by Bratteli diagrams and inductive
limits) and lead to ``affine"  Hecke algebras [71]. All these new degrees of
freedom permit the construction of local lattice theories (like RSOS) via
Hamiltonians or transfer matrices. Although the number of possibilities is
``myriotic", one expects to recover (through scaling limits) only those
continuous new degrees of freedom which are permitted by the principles of
algebraic QFT (in case the lattice models have second order phase-transition
points).

One would hope that algebraic QFT is capable of combining all those different
discoveries, including those in conformal field theory (independent of whether
they had been made by ``truffel hogs" or supersophisticated mathematical
physicists) under one roof: the classification theory of low-dimensional QFT's
and the explicit construction of their free plektonic representations.
Despite some nice progress in the understanding of momentum space behaviour of
plektons through the powerful formalism of scattering theory, I think that
there is one important corner stone missing: quantum symmetry.

Internal symmetry in ordinary quantum physics amounts to  representation theory of
compact Lie groups. This is at least our picture ever since Heisenberg
introduced isospin. It took a very long time  and, physically as well as
mathematically, sophisticated work, to see that group symmetry is a consequence
of quantum field theoretical permutation group statistics which in turn
(at least for $d=4$) results from Einstein causality of observables and
localization of states. The DHR [1] work and its DR [72] completion, despite its 
monumental conceptual enrichment, has not found widespread appreciation,
and the reason for this is obvious. They proved that what one can derive from
intrinsic quantum principles is just the same as  what one had known
all along from quantization of classical Lagrangians, at least as far as
internal symmetry in $d=4$ is concerned. 

There exists a more nonlocal looking but physically equivalent
description of QFT where the charge-carrying fields are not multicomponent
bosons or fermions but rather
 ``parafermions"\footnote{We use the word in the Green, Messiah, Greenberg, 
and DHR sense, but not in Kadanoff's.}
 of some finite order  equal to the hight of the statistics Young tableaux which
obey R-matrix commutation relations with trivial monodromy $R^2=1$[73]. They
live in a smaller Hilbert space without the group theoretic multiplicities.
They lead to the same scattering observables and are built on the same
observable-algebra, but the amplification factors
in cross sections (which in the group theoretic description enter through
traces over Casimir-operators) originate for parafermions from the unusual
inner products for scattering states (given by a Markov trace on the 
permutation group).
The standard description in terms of multicomponent fermions and bosons is the
better  one: it is the only one which allows a Lagrangian formulation i.e. a
description by quantization, whereas the para-statistics-description is too 
non-commutative for that purpose. Physically more important is that the standard description
is required if it comes to the important issue of symmetry-breaking
(an issue which has nothing to do with quantization).

Around 87/88 the question of what symmetry concept is behind braid group
statistics arose, and it was clear that it cannot be  a symmetry 
arising from a compact group .

There was an obvious formal relation between the ``$q$-dimensions" of 
quantum groups
and the DHR-Jones-Wenzl quantization via tracial Markov states on the
Jones-Temperly-Lieb algebras or on {\bf C}$B_\infty$ braid group algebras
respectively (in the original DHR approach on {\bf C}$S_\infty$).
So it was very tempting indeed to interpret quantum groups as being the (dual)
symmetry behind braid group statistics. In a spirit of excitement I was
already on my way (in 87/88) to write up notes, when Rehren (to whom I am
very grateful) pointed out some flaws of quantum groups in relation to the
superselection theory. Some of these (the more mathematical ones) were later
overcome in the work of Mack and Schomerus [74] by the introduction of ``weak
quasi"-Hopf algebras. But the nonuniqueness of their construction (in the DHR
theory the symmetry was unique) as well as the somewhat asymmetric looking
treatment of antiparticles (i.e. the conjugation in weak quasi-Hopf algebras)
dampened my enthusiasm somewhat. The asymmetry 
 property is of course shared with
our previous ``exchange algebras", but they at least were unique. 
Their algebraic relations (as well as ours)
are incomplete i.e. a distinction between
non-overlapping and overlapping localization has to be made.
	
So I began
to ask the question: why do the well-understood unique exchange algebras in
the DHR framework (where they amount to parastatistics with R-matrix 
commutation relations) are less than acceptable as compared to the DR compact
Lie group (i.e. the standard) tensorial description?

One answer was a very pragmatic one. Old theorems like the previously mentioned
Jost-Schroer-Pohlmeyer [2] theorem which relates e.g. the algebraic assumption
of free field equations (as defining an ideal within a formal field algebra,
the Borchers algebra) to the free $n$-point correlation function with its
Wick-combinatorics, such old theorems run into trouble with parastatistics
fields. In that case one does not start with tensorial fermions or bosons
(the para-fields are localizable i.e. have R-matrix commutation relations, but
they are not local) and one is not supposed to end with the Wick-combinatorics
either: parastatistics free fields have a much more complicated combinatorics.
All known formalism, including the nonrelativistic generalized Hartree-Fock
methods, rely on local tensorial fields.
In addition there is the important issue of symmetry breaking  and one does
not have the slightest idea of how to do this with Mack-Schomerus fields. 

In
this complex situation I went back and looked again at our previous construction
of the universal observable algebra ${\cal A}_{uni}$  which we elaborated in algebraic
conformal QFT (following a previous more abstract proposition by Fredenhagen).
The physical idea which led to its construction was very simple. Whereas 
in the old DHR theory with permutation group statistics, the net was
naturally directed towards Minkowski-space infinity and therefore you could
construct outer endomorphisms (charged sectors) as a limit of inner charge
transport by just disposing any unwanted anti-charge into the big waste
basket at infinity [1], this was not possible for non-directed nets as e.g.
conformal nets. 

In that case, the only natural globalization was that 
through free, amalgamated (over all local relations) $C^*$-algebras, namely
${\cal A}_{uni}$. The key physical idea was that every localized endomorphism
$\rho$ (or its sector-equivalence class) gives rise to a global 
selfintertwiner [FRSII, ref.58]
$V_\rho$ (charge-transport ``once-around") 
described by a unitary operator within
${\cal A}_{uni}$, and that its use in an intertwining chain:\\
vacuum $\mathop{\longrightarrow}\limits^{split}$ charge-anticharge
$\mathop{\longrightarrow}\limits^{selfintertwining}_{of charge}$
charge-anticharge $\mathop{\longrightarrow}\limits^{fusion} $ vacuum\\
leads to invariant global charges and (by spectral decomposition) to
invariant projectors inside ${\cal A}_{uni}$, which project onto the various
central components of ${\cal A}_{uni}$. It immediately led to a very profound
understanding of Verlindes observations on characters of Virasoro or 
W-algebras and in addition one obtained that the globalized exchange algebra
is really living on a helix (the covering of $S^1$): the true braid group
relations are those of $B_\infty$ on a cylinder and not on the plane
(or on a Riemann sphere). The covering is finite (a certain power of the
self-intertwiner is trivial) if the model is ``rational", but otherwise
our results were of a completely general model-independent nature (it can be
easily transferred to 3-d theories). 

Recently I realized that there is much
more in ${\cal A}_{uni}$. If one uses these intertwiners $V_\alpha$ inside more
complicated splitting chains according to:

$$\rho \mathop{\longrightarrow}\limits^{split} \quad \alpha \circ \beta\quad
\mathop{\longrightarrow}\limits^{\alpha-}_{selfintertwining} 
\quad \alpha \circ \beta \quad
\mathop{\longrightarrow}\limits^{fusion}\quad  \rho'$$

\qquad (where $\alpha$ and $\beta$ are two irreducible intermediate sectors)\\
then the result is that within ${\cal A}_{uni}$ one can construct
intertwiners between sectors $\rho$ and $\rho'$ (which only
depend on the $\alpha,\beta$ equivalence classes). In such way one easily
obtains charge-creating intertwiners which e.g. connect the vacuum with some
other irreducible sector. There is, however, a glitsch to this: the intertwining
is a $C^*$-algebra property which is lost in its vacuum representation. 
It only becomes activated in higher representations i.e. in the presence
of other charges. For this reason I called this mechanism  
``charge-polarization"
symmetry [75]. According to my best knowledge it is also a mathematically new
phenomenon, i.e. not part of known representation theory. It can be traced
back to the ``freeness" aspect of the ${\cal A}_{uni}$ $C^*$-algebra.
For ``rational" theories (i.e. a finite number of sectors) one can see that
the subalgebra of such intertwiners which one obtains by fixing a basis
of endomorphisms is finitely generated with maximally $N^3$ 
generators in the case of $N$ sectors; the
third power is coming about because for each sector there are $N$ global
$V_\rho^n$-intertwiners). This algebra leads to a refinement of central projectors
to non-central ones: just multiply the central projectors with the various
projectors obtained by spectral decomposition of $V_\rho^n$, using the finite
spectrum (monodromy phases) which those operators have on the generating
fusion intertwiners. 

For the first time, one found a structure which is characteristic
for the physics of plektons, it is blind against fermions and bosons.
In the DHR theory, global charge operators can only be found in specific
representations (as weak limits) and never in the global so-called
``quasilocal" observable algebra with that big waste disposal at infinity.

If you ask what geometric structure this global intertwiners algebra
represents, my tentative answer is that the polarization symmetry has a very 
intimate connection to some sort of ``universal mapping class group"-algebra
amalgamated over ``something". Although I have no proof for this (this notion
is presently not even well-defined), I have done numerous graphical checks
using the physicist's results on mapping class group representations in
conformal field theory or combinatorics. 

I am convinced that the attempts to define and classify conformal QFT in
terms of objects on Riemann surfaces is like doing things upside-down 
(at least from a quantum physical standpoint). 

There are two ways in which
Riemann surfaces may be related  to chiral conformal QFT. On the one hand
they may serve as a mnemotechnical device to keep track of the ternary fusion
structure and the combinatories. On the other hand, Riemann surfaces could
also serve as the generalized Bargman-Hall-Wightman domains of analytically
continued correlation functions (but never as the localization space on which
fields ``live") which then would have complicated automorphic properties.
These complicated automorphic properties could result by averaging the 
original real-time correlation function with the help of Fuchsian groups
(one formally maintains the positivity on light rays). Wheras for $g=1$ this
averaging can be controlled  [76] and dumped into the states 
(which thereby become the well-known $L_0$-temperature states),
 there is  no understanding of these formal
manipulations for $g\ge2$. In this case, also the algebra seem to suffer
a radical change, and it is  unknown, what this means in terms of algebras 
and states.

On a very formal level one makes, however, the curious observation that the
original net indexing (in this case by intervalls on the light ray) looses
its meaning in terms of Einstein causality and one finds a kind of super
Reeh-Schlieder situation: what used to be the old net members now become dense
in the total algebra (as a result of the Fuchsian
\footnote{The Fuchsian groups for higher genus are discrete subgroups
of the symmetry group of the vacuum state.}
 averaging). This is one
possible scenario of how Einstein causality may be outmaneuvred: there is no
global notion of space-like and hence no necessity to find compatible
(commuting) measurements in causally disjoint regions.
The standard causality could then only survive in the infinitesimal.

Similarly the 3-mf. invariants cannot be related to the ``living space"
of fields (TFT cannot be used to extract information on localization), they
rather belong to that ill-understood ``twilight zone" of external/internal
symmetry.

In discussing the aforementioned global structures of ${\cal A}_{uni}$ with
Karowski, I also became aware of the fact that the selfintertwiners $V_\rho$
correspond to lines around handles in their [77]
 (together with Schrader) formulation of invariants of graphs on surfaces.
 This gives a geometrical 
interpretation of the strange ``flip relations" which these global intertwiners
have with local ones. I also learned from him that by extending somewhat the
Pasquier lattice construction of generalized ABF-like models, 
one may be able to see some of the above charge
creating properties already on a finite lattice [78]. 

The crucial test for
physical usefulness of quantum symmetry based on the charge polarization idea,
however, has not yet been carried out. In analogy with the parafermions
versus compact Lie  group fermions of before, I think that it should play
a crucial role in converting the  momentum-space description of plektons into
localizable free plektonic $x$-space fields with a ``generalized Fock space".

The relation of nonlocal fields to localizable fields has an interesting
analogue in the Bethe-Ansatz approach to integrable lattice models. There
the so-called B-field, which  related to the pseudo-vacuum and the generated
pseudo-excitations, gives rise to the physical vacuum and localizable physical
vector states whose relation to localizable charge carrying fields
 is, however, technically too difficult in order
to be explored analytically (a typical difficulty of lattice physics, where
the start is always conceptually easy, but life becomes very tough later on).

So what I am presently ``dreaming" about is that QFT ``liberation" of the
Bethe-Ansatz idea, which Rehren and I verbally already mentioned in our
``Einstein causality and Artin braid" paper. In any case, I do not share
Fredenhagen's optimism (expressed at  the Paris sattelite conference)
of expecting to get the full description of the
$x$-space localized plektons including their $n$-point correlation functions
within the extended exchange algebra
framework (without using ``quantum symmetry").

There is another potential use of these ideas, this time of a more mathematical
nature. The present understanding of invariants of 3-manifolds either by
functional integral or combinatorial recipes or even (via the $\rho_{reg}$-idea
mentioned before) with algebraic QFT leaves a lot to be desired from a physical-philosophical standpoint.

Analogous to the Coleman-Mandula-O'Raifeartaigh [66] 
No-Go theorem concerning the marriage between
space-time symmetries and inner symmetries in the old theory, there should be
a compelling reason, why now, in the context of braid-group statistics, these
two concepts are inexorably linked. This is of course an issue which was
already there ever since Jones derived knot invariants. 

Algebraic QFT and
the Jones subfactor  theory  gives a very good mathematical control and describes this
phenomenon very well, but, at least presently, it does not really ``explain"
 it.

One would hope, that a quantum symmetry concept, which is sufficiently
different from the old one, is able to achieve that. It is interesting that
Popa, [79] in some recent work, expresses the hope that some ``free product" construction
for operator algebras may help in unravelling some yet hidden geometrical structures 
in the Jones theory.
I think that algebraic QFT has enough conceptual richness to solve such
problems and I hope that in this way we may turn around the (up to now) rather
one-sided subfactor -- algebraic QFT connection. 

A real high-energy particle
physicist, in taking notice of these developments, may think that these are
isolated ideas, perhaps applicable to condensed matter physics, but never of any
relevance to 4-dimensional gauge theory and quark physics. I do not share
such a pessimistic view. 

On the contrary, I think that after a much better
understanding of low-dimensional QFT has been achieved, mathematical physicists
will revisit gauge theory and the not yet understood problem of quark
confinement, and finally realize that the relation of e.g. QCD to a new theory
(not built on quantization) will be similar to the relation of the quasiclassical
Bohr atomic model to the full Q.M.. Comparing the successful low-dimensional
concepts with the gauge theory based on the quantization of fibre bundles,
one gets the impression that QCD is much too classical for a fundamental
quantum theory. 

Why (apart from an associated short distance scaling theory
as mentioned before) should quarks in their role as physical spectrum
generating objects be describable as Wigner particles (i.e. have a 
welldefined spin and mass assignement as it is characteristic for the class of localizable
finite energy states)? In what sense can QCD lead to the problematization
of the concept of ``magnetic field" 
with an ensuing ``self duality", generalizing that of plektons
(on the same profound quantum level as
the recent insight into low-dimensional QFT arose from a problematization of
``charges")?

 I would like to think of quarks as being basically physical quantum states
\footnote{I do not think that Lagrangian gauge theories like QCD would allow
to do that (see chapter 2, page 15). For critical observations on the quantization
of nonabelian gauge theories by functional integrals, see [80].}
whose main ``unphysical" property is that they have infinite energy (and very weak
localization properties). According to our present best knowledge, semiinfinite
space like cone localization either corresponds to spectral mass gaps
(in that case there is unitary equivalence to arbitrarily thin i.e. Mandelstam string-like
localization) or to infraparticles (where the shape of the photon cloud 
determines the localization size). 

Therefore infinite energy quantum states, which are
even more remote from Wigner particles than infraparticles, are likely to
admit a localization which is not better than that around space-like surfaces.
This then could again lead to a rich geometric structure (space-like surfaces
allow for a generalized braiding) and a possible generalization of statistics.

Zamolodchikov's invention of the tetrahedral equations (with promising
attempts of Baxter et al. in the direction of 3-dimensional integrable
lattice-model building), together with the necessity to find a quantum 
problematization of the classical notion of ``magnetic field" (as it was
successfully accomplied for ``charges") are ominous and hopeful signs.
But presently one lacks a physical principle leading to such strange
states and fields.

After the discovery of Q.M., almost no conceptual breakthrough has been 
made with just one attempt. A good illustration is conformal QFT, which required
three attempts (well separated in time) in order to reach its present 
perfection. One would be inclined to believe that deep notions on the QFT
 S-matrix connections (like e.g. the Borchers equivalence classes of local
fields) will have their comeback on a higher level of conceptual sophistication.
Since the present generation of theoretical physicists has a very superficial knowledge
of relativistic scattering theory (some none at all, or only through the
string theoretic carricature of an S-matrix) this will probably not happen
in the forseeable future, unless the older generation does transmit the
knowledge of these incomplete old discoveries through lectures and in courses
on QFT. If, however, physics looses its historical connection and in every
decade new ahistorical inventions with new terminologies are proposed, or if
mathematical physics is just used for spare parts in pure mathematics, then
our future is rather doubtful. In that case I also do not understand why
society should financially support our activities.

\newpage
\noindent{\LARGE\bf References and Additional Comments}

\begin{enumerate}
\item[1]  R. Haag, ``Local Quantum Physics", Springer Verlag, Berlin (1992)
\item[2]   R.F. Streater and A.S. Wightman, ``PCT, Spin and Statistics,
and All That", Benjamin, New York (1994).\\
R. Jost, ``General Theory of Quantized Fields", Americ. Math. Soc. Publication, 1963.
\item[3] R. Haag and B. Schroer, ``Postulates of Quantum Field Theory",
J.Math. Phys. {\bf3}, 248 (1992). 
\item[4] G.C. Hegerfeldt, ``Causality Problem for Fermi's Two-Atom System",
Phys. Rev. Lett.{\bf72}, 596 (1994).
\item[5] D. Buchholz and J. Yngvason, ``There are no Causality Problems of
Fermi's Two Atom System", Phys. Rev. Lett.{\bf 73}, 613 (1994).
\item[6] J. Langerhole and B. Schroer, ``Can Current Operators Determine a
Complete Theory?", Commun. Math. Phys. {\bf4}, 123 (1967).
\item[7] K.H. Rehren, ``News from the Virasoro Algebra, DESY preprint 93/115.
\item[8] J. Glimm and A. Jaffe, ``Quantum Physics, a Functional Integral
Point of View", Springer 1987.
\item[9] B. Schroer, ``Infrateilchen in der Quantenfeldtheorie", Fortschr.
Phys. {\bf 173}, 1527 (1963).\\
D. Buchholz, ``On the Manifestations of Particles", proceedings of the
``Workshop on Mathematical Physics Towards the 21st Century", Ed. R.N. Sen
and A. Gersten, Beer-Sheva, Israel 1993, Ben-Gurion University of the
Negev Press 1994.
\item[10] B. Klaiber, Lectures in Theoretical Physics, Boulder 1967, p.141,
Gordon and Breach, New York, 1968.
\item[11] H. Lehmann and J. Stehr, ``The Bose Field Structure Associated with
a Free Massive Dirac Field in One Space Dimension", DESY report 76/29, June 1976,
unpublished.
\item[12] B. Schroer and T.T. Truong, ``Equivalence of the Sine-Gordon and
Thirring Models and Cumulative Mass Effects", Phys. Rev.{\bf D15}, 1684 (1977).
\item[13] B. Schroer, ``Operator Approach to Conformal Invariant Quantum Field
Theories and Related Problems", Nucl. Phys. {\bf B295}[FS21] 586(1988), page 594.
\item[14] A.S. Wightman, ``Proceedings of the Fifth Coral Gables Conference
of Symmetry Principles at High Energy", University of Miami, 1968, ed. Gudehus et al..
\item[15] R. Seiler, B. Schroer, and J.A. Swieca, ``Problems of Stability for
Quantum Fields in External Time-Dependent Potentials", Phys. Rev. {D2}, 2927 (1970).\\
A S-Matrix construction for higher spin particles 
with the main emphasis on covariance where problems of stability and
causality were not yet considered, has been given by Weinberg:\\
S. Weinberg, ``Feynman Rules for Any Spin", Phys. Rev. {\bf 133B}, 1318 (1964).
\item[16] A. Pressley and G. Segal, ``Loop Groups", Oxford University Press (1986).
\item[17] S.N.M. Ruijsenaars, ``Quantum theory of relativistic  charged particles 
in external fields", Thesis, Leiden, Netherland (1976).
\item[18] J. B\"ockenhauer, ``Localized Endomorphisms of the Chiral Ising Model",
DESY preprint 1994.
\item[19] J.A. Swieca, ``Goldstone's theorem and related topics", Carg\'es
Lecturs in Physics, Vol.{\bf 4}, p.215 (1970).
\item[20] J.A. Swieca, ``Charge screening and mass spectrum", Phys. Rev. {\bf D13},
312 (1976).
\item[21] E.C. Marino and J.A. Swieca, ``Order, Disorder and Generalized Statistics",
Nucl. Phys. {\bf B170} [FS1] 175 (1980) page 181.
\item[22] B. Schroer, ``A Theory of Critical phenomena based on the Normal
Product Algorithm", Phys. Rev. {\bf B8}, 4200 (1973).
\item[23] K.D. Rothe and B. Schroer, ``Do Quark-Correlation Functions exist
on Confining Gauge Theories?", Nucl. Phys. {\bf B172}, 383 (1980).
\item[24] E. Abdalla, M. Cristina B. Abdalla and K.D. Rothe, ``Non-perturbative
methods in 2-Dimensional Quantum Field Theory", World Scientific Publishing Co. 1991.\\
The chapter I-XII are a useful source of information on the ``first generation"
work on two-dimensional solvable models.\\
All the calculations in the book which illustrate ``screening versus confinement"
were done with the correct picture in mind, (Confinement = surpression of fractional
flavour which one adds to the model as a testing charge) as in the
Kurak-Schroer-Swieca paper quoted therein.\\
 Nevertheless at the end of the conclusions in 10.4 the old aborted ideas of
linking this distinction with poles 
(and scattering states) again are looming through. To be sure, one can discuss
the issue of ``free charge versus screeing versus confinement" concretely with
correlation functions, but than one has to use the criteria given in:\\
K. Fredenhagen, ``Particle Structure of Gauge Theories",
published in Erice school on Mathematical Physics 1985, p.265.\\
\item[25] G. Mack and K. Symanzik, ``Currents, stress tensor and generalized 
unitarily in conformal invariant quantum field theory", Commun.Math.Phys.
{\bf 27}, 247 (1972).
\item[26] G. Mack, ``Group theoretic approach to conformal invariant quantum
field theory", J.de Physique (Paris) {\bf 34} C1 (suppl. no.10), 99 (1973).
\item[27] R. Seiler,  M. Horta\c csu and B. Schroer, ``Conformal Symmetry
and Reverberations", Phys. Rev. {\bf D5}, 2519 (1972).
\item[28] B. Schroer and J.A. Swieca, ``Conformal transformations for quantized
fields", Phys. Rev. {\bf D10}, 480 (1974).\\
B. Schroer, J.A. Swieca and A.H. V\"olkel, ``Global operator expansions in
conformally invariant relativistic quantum field theory", Phys. Rev. {\bf D11}, 11
(1975).\\
The conformal QFT  in most of the post 1980 papers is presented in the following
form: one contemplates the existence of a
chiral field $\varphi(z)$ with $z$ not just on the circle or the real line but being
a variable in the complex plane and imagines holomorphic transformation
properties in that variable. 

It is well-known that this in not even true for
free fields (the positive frequency part can be extended to the upper half-plane
and the negative frequency part is analytic in the lower half-plane). Of course
there are vector-states and correlation function (but not algebras and operators!)
which have analytic properties, namly those studied by Bargman, Hall and Wightman [2].
But the global conformal transformation properties in the $z$-variable of
analytically extended correlation functions  in most of the literature are incorrect:
they simply contradict the subtle analytic branching properties of those
functions (e.g. the relevance of $\widetilde{SL(2,R)_c}$ versus $SL(2,${\bf C})).\\
Fortunately people use calculational recipes which allow to bypass the confrontation
of what they say with what they think they mean. But in conceptual or structural
problems as those of the aborted paper [4] or the resolution of the Einstein
causality paradoxon via central projection of the present paper,
 this ``mild" muddyness would lead to disaster.\\

\item[29] B. Schroer, ``A Trip to Scalingland", Brazilian Symposium on
Theoretical Physics", Rio de Janeiro, January 1994,
Vol.I , ed. Erasmo Ferreira, Livros T\'ecnicos e Cientificos Editora S.A..
\item[30] R. Dashen and Y. Frishman, ``Thirring Model with U(n) Symmetry",
Institute for Advanced Study, Princeton report 1973 (unpublished).\\
P.K. Mitter and P.H. Weisz, ``Asymptotic Scale Invariance in a Massive Thirring
Model with U(n) Symmetry", Phys. Rev. {\bf D8}, 4410 (1973).
\item[31] N.K. Nielsen, K.D. Rothe and B. Schroer, ``Fermionic Green Function
and Functional Determinant in QCD$_2$", Nucl. Phys. {\bf B160}, 330 (1979).
\item[32] K. Gawedzki, ``Topological Actions in Two-Dimensional Quantum
Field Theories", in Nonperturbative Quantum Field Theory NATO ASI Series B,
Physics 1988.
\item[33] K.D. Rothe and B. Schroer, ``Liberation of Exotic States in Two-Dimensional
Abelian Gauge Theories", Nucl.Phys. {\bf B185}, 429 (1981).\\ \\
The idea to look for the liberation of kinks in two-dimensional gauge theories,
we got from Coleman's semiclassical gauge-theoretic picture of his ``half-space
kinks" [see Coleman's Erice lecturs]. 
\item[34] K. Fredenhagen, ``Generalization of the Theory of Superselection Sectors",
in: The Algebraic Theory of Superselection Sectors, Introduction and Recent
Results, Ed. D. Kastler, World Scientific Publishing Co. 1990. \\ \\
This work is closely related to the Buchholz-Fredenhagen paper quoted therein.
\item[35] B. Schroer and J.A. Swieca, ``Spin and Statistics of Quantum Kinks",
Nucl.Phys. {\bf B121}, 505 (1977).
\item[36] J.H. Lowenstein and J.A. Swieca, ``Quantum Electrodynamics in Two
Dimensions", Annals of Physics {\bf 68}, 172 (1971).
\item[37] N.K. Nielsen and B. Schroer, ``Topological Flucturations and Breaking
of Chiral Symmetry in Gauge Theories involving massless Fermions", Nucl. Phys.
{\bf B120}, 62 (1977).\\
N.K. Nielsen and B. Schroer, ``Saturation of Gauge-Invariant Schwinger Model
Correlation Functions by Instanton", Phys. Lett. {\bf 66B}, 373 (1977).
\item[38] M. Horta\c csu, K.D. Rothe and B. Schroer ``Still more about the
fermions determinent in two-dimensional QED", Phys.Rev. {\bf D22}, 3145 (1980).
\item[39] N.K. Nielsen and B. Schroer, ``Axial Anomaly and Atiyah-Singer Theorem",
Nucl.Phys. {\bf B127}, 493 (1977).\\
N.K. Nielsen, H. R\"omer and B. Schroer, ``Classical Anomaly and Local Version
of the Atiyah-Singer Theorem", Phys. Lett. {\bf 70B}, 445 (1977).
\item[40] An example is the derivation of the nonlocal boundary conditions of the
topologist for the Dirac equation from C-invariance and chiral invariance
(apart from zero modes), M. Horta\c csu,
K.D. Rothe and B. Schroer, ``Zero-Energy Eigenstates for the Dirac Boundary
Problem", Nucl. Phys. {\bf B171}, 530 (1980).
\item[41] R.J. Crewther, ``Chiral Properties of Quantum Chromodynamics", in
Field Theoretical Methods in Particle Physics, 1979, Nato Advanced Study
Institutes Series, Series B: Physics, Vol. 55, ed. W. R\"uhl.
\item[42] V.E. Korepin and L.D. Faddeev, ``Quantization of Solitons",
Theor. Math. Phys. {\bf 25}, 1039 (1975).
\item[43] B. Schroer, T.T. Truong and P. Weisz, ``Towards an Explicit Construction
fo the Sine-Gordon Field Theory", Phys. Lett. {\bf 63B}, 422 (1976).
\item[44] A.B. Zamolodchikov, `'Exact Two-Particle S-Matrix of Quantum Sine-Gordon
Solitons" Commun. Math. Phys. {\bf 55}, 183 (1977).
\item[45] M. Karowski, H.J. Thun, T.T. Truong and P. Weisz, ``On the Uniqueness
of a purely elastic S-Matric in (1+1) Dimensions", Phys.Lett. {\bf 67B}, 321 (1977).
\item[46] B. Berg, M. Karowski, V. Kurak and P. Weisz, ``Factorized U(n)
Symmetric S-Matrices in Two Dimensions", Nucl. Phys. {\bf B134}, 125 (1978).
\item[47] V. Kurak and J.A. Swieca, ``Antiparticles as Bound States of Particles
in the Factorized S-Matrix Framework", Phys. Lett. {\bf 82B}, 289 (1979).\\
see also R. K\"oberle and J.A. Swieca, ``Factorizable Z(N) Models", Phys. Lett.
{\bf 86B}, 209 (1979).\\
R. K\"oberle, V. Kurak and J.A. Swieca, ``Scattering Theory and 1/N Expansion
in the Chiral Gross-Neven Model", Phys.Rev. {\bf D20}, No.4, 897 (1979).
\item[48] K.D. Rothe and J.A. Swieca, ``Fractional Winding Numbers and the
U(1) Problem", Nucl. Phys. {\bf B168}, 454 (1980).
\item[49] M. Karowski, V. Kurak and B. Schroer, ``Confinement in Two-Dimensional
Models with Factorization", Phys. Lett. {\bf 81B}, 200 (1979).
\item[50] M. Karowski, ``On The Bound State Problem in 1+1 Dimensional Field
Theories", Nucl. Phys. {\bf B153}, 244 (1979).
\item[51] P.P. Kulish, N.Yu. Reshetikhin and E.K. Sklyanin,
``Yang-Baxter Equations and Representation Theory I", Letters in Math.Phys.
{\bf 5}, 393 (1981).
\item[52] M. Karowski and H.J. Thun, ``Complete S-Matrix of the O(2N)
Gross-Neveu Model", Nucl. Phys. {\bf B190} [FS3], 61 (1981).
\item[53] J. Fr\"ohlich, ``Quantum Sine-Gordon Equation and Quantum Solitons
in Two Space-Time Dimensions" in ``Renormalization Theory", Nato Advanced
Study Institute Series, Series C, Vol. 23, Erice 1975.
\item[54] E.C. Marino and J.A. Swieca, ``Order, Disorder and Generalized 
Statistics", Nucl. Phys. {\bf B170}, 170 (1980).\\
E.C. Marino, B. Schroer and J. A. Swieca, ``Euclidean Functional Integral
Approach for Disorder Variables and Kinks", Nucl. Phys. {\bf B214}, 414 (1983).\\
B. Schroer, ``Functional Integrals for Order-Disorder Variables in Terms of
Aharonov-Bohm Strings", Nucl. Phys. {\bf B120} [FSG], 103.\\
One interesting result which I never published, was that the 2-d $\phi^4$ kink
in the euclidean approach leads to a two-sheeted solution for the relativistic
kink-kink-field 3-point function (which obeys the same nonlinear differential
equation as the spontaneous vacuum expectation value). I was not able to find
an analytic solution, but through a correspondence with C. Taubes, I learned
how to prove the existence of a solution which approaches $\pm1$ at infinity
on the first resp. second sheet, and which branches along a cut going from the
euclidean position of one kink to the other. This 3-point function was to be
used in a relativistic perturbation approach to kinks in analogy to the vacuum
1-point function in the perturbation theory of the broken symmetry phase.\\
The notes can be made available to anybody who wants to work
on this problem.
\item[55] J. Fr\"ohlich and P.A. Marchetti, ``Superselection Sectors in Quantum,
Field Models: Kinks in $\phi^4_2$ and Charged States in Lattice (Q.E.D.)$_4$"
in ``The Algebraic Theory of Superselection Sectors. Introduction and Recent
Results" Ed. D. Kastler, World Scientific 1990.
\item[56] See Fr\"ohlich's and my contributions in ``Nonperturbative Quantum
Field Theory" (Carg\`ese 1987), G, 't Hooft et al. (eds.) New York:
Plenum Press 1988.
\item[57] K.H. Rehren and B. Schroer, ``Exchange Algebra and Ising n-point
Functions", Phys.Lett. {\bf 198}, 84 (1987).
\item[58] K. Fredenhagen, K.-H. Rehren and B. Schroer, ``Superselection sectors
with braid group statistics I", Commun. Math. Phys. {\bf 125}, 201 (1989).\\
K. Fredenhagen, K.-H. Rehren and B. Schroer, ``Superselection sectors with
braid group statistics and exchange algebras II", Rev.Math.Phys. Special ussue,
113 (1992).
\item[59] A. Klein and L.J. Landau, ``Stochastic Processes Associated with KMS
States", Journal of Functional Analysis {\bf 42}, 368 (1981).
\item[60] F. Nill, ``A constructive quantum field theoretic approach to Chern-Simons
theory", International Journal of Mod.Phys. {\bf B6}, 2159 (1992).
\item[61] Narnhofer and W. Thirring, ``Covariant QED without Indefinite Metric",
Rev. in Math. Phys., Special Issue (1992).\\
F. Acerbi, G. Morchio and F. Strocchi, ``Theta Vacua, Charge Confinement and
Charged Sectors from Nonregular Representations of CCR Algebras",
Letters in Math. Phys. {\bf 27}, 1 (1993).
\item[62] S. Doplicher, K. Fredenhagen and J.E. Roberts, ``Spacetime Quantization
induced by Classical Gravity", to be published in Physics Letters.
\item[63] D. Buchholz and R. Verch, ``Scaling Algebras and Renormalization Group
in Algebra Quantum Field Theory", DESY preprint, July 1994.
\item[64] K. Fredenhagen, M. Gaberdiel and S.M. R\"uger,
 ``Scattering states of Plektons in 2+1 Dimensional Quantum Field Theory",
preprint DAMPT 94/90.
\item[65] B. Schroer, ``Modular Theory and Symmetry in QFT", proceedings of the
``Workshop on Mathematical Physics Towards the 21st Century", Ed. R.N. Sen
and A. Gersten, Beer-Sheva, Israel 1993, Gen-Gurion University of teh Negev
Press 1994.
\item[66] S. Coleman and J. Mandula, ``All possible symmetries of the S-Matrix",
Phys.Rev. {\bf 159}, 1251 (1990).\\
L. O'Raifeartaigh, ``Mass differences and Lie algebras of finite order",
Rev.Mod.Phys. {\bf 42}, 381 (1970).
\item[67] F. Wilszek, ``Quantum Mechanics of Fractional-Spin Particles",
Phys.Rev.Lett. {\bf 49}m No. 14, 957 (1982).
\item[68] M. Leinaas and J. Myrheim, Nuovo Cimento {\bf 37B}, 1 (1977).\\
J. Mund and R. Schrader, ``Hilbert Spaces for Nonrelativistic and Relativistic
'Free' Plektons", to appear in the Proceedings of the Conference ``Advances
in Dynamical System and Quantum Physics", Capri (Italy) May 1993.
\item[69] Lattice system allow a very similar conceptual framework to QFT.
The main distinction  is that the localization properties become more complicated
and as a result of this, the derivation of particle states and their scattering
theory requires much more work.\\
J.C.A. Barata, ``S-Matrix Elements in ``Euclidean Lattice Theories", Reviews
in Mathematical Physics, Vol.6, No.3, 497 (1994).
\item[70] V. Pasquier, ``Etiology of IRF Models", Commun. Math. Phys. {\bf 118},
355 (1988).
\item[71] This is presumably related to a new class of solvable lattice
models in which the lattice is the surface of a cylinder with the inside
consisting of pure ``combinatorial stuff".\\
M. Karowski and A. Zapletal, ``Quantum-group-invariant integrable n-state
vertex models with periodic boundary conditions", Nucl. Phys. B419 [FS] (1994).

\item[72] S. Doplicher and J.E. Roberts, ``Why there is a field algebra with
a compact gauge group describing the superselection structure in particle
physics", Commun.Math.Phys. {\bf 131}, 51 (1990).
\item[73] B. Schroer, ``Algebraic QFT as a Framework for Classification and 
Model-Building" in ``The Algebraic Theory of Superselection Sectors.
Introduction and Recent Results", Ed. D. Kastler, World Scientific Publishing Co.
(1990).
\item[74] G. Mack and V. Schomerus, Quasi Hopf Quantum Symmetry in Quantum Theory,
Nucl.Phys. {\bf B370} 185 (1992).
\item[75] B. Schroer, ``Universal observable Algebras and Polarization Symmetry",
FU Berlin preprint, October 1994.
\item[76] B. Schroer, ``Quantum Field Theory on Riemann Surfaces and the
Unitarity Problem", Phys. Lett. {\bf 199}, 183 (1987).
\item[77] M. Karowski and R. Schrader, ``A combinatorial approach to
topological quantum field theories and invariants of graphs",
Commun.Math.Phys. {\bf 151}, 355 (1993).
\item[78] M. Karowski and R. Schrader, ``A lattice model of local algebra
of observables and fields with braidgroup statistics", in preparation.
\item[79] S. Popa, ``Markov traces on universal Jones algebras and subfactors
of finite indes", UCLA preprint 1992.

\item[80] M.S. Marinov, ``Quantization of Field Theory with Nontrivial Geometry",
proceedings of the ``Workshop on Mathematical Physics Towards the 21st 
Century", Ed. R.N. Sen and A. Gersten, 
Beer-Sheva, Israel 1993, Gen-Gurion University of the Negev Press 1994. \\
According to my best knowledge this is the most poignant criticism
within the standard setting of QFT (e.i. not using methods of algebraic QFT).

\end{enumerate}

\end{document}